\documentclass[11pt]{article}
\usepackage[utf8]{inputenc}
\usepackage{amsmath, amssymb, amsthm,amsfonts}
\usepackage{graphicx}
\usepackage{subcaption}
\usepackage{float}
\usepackage{geometry}
\usepackage{xcolor}
\usepackage[title]{appendix} 
\newtheorem{proposition}{Proposition}

\title{Properties of Adjoint Solutions of the Full-potential Equations for Two-Dimensional Subcritical Flow}
\author{Carlos Lozano and Jorge Ponsin  \\
Theoretical and Computational Aerodynamics  \\
National Institute of Aerospace Technology (INTA), Spain  \\
*Corresponding author: lozanorc@inta.es }
\date{}

\begin{document}

\maketitle

\begin{abstract}
The adjoint full-potential equations are studied for two-dimensional (2D) steady subcritical
flows. In contrast with the incompressible case, explicit closed-form solutions are generally
not available in the compressible setting, so the emphasis is placed here on the underlying structure. Using the Green's-function approach and the relation between the adjoint
full-potential and compressible adjoint Euler equations, we identify the adjoint potential
and stream function with linear combinations of the Euler adjoint variables associated with
point mass and vorticity sources. For lift-based cost functions, the corresponding adjoint
solutions contain two unknown functions that encode the effect of perturbations to
the Kutta condition. We show that these functions obey the linearized full-potential
equations, are linked by generalized Cauchy-Riemann equations, and reduce in the
incompressible limit to the Poisson kernel of the Laplacian on the exterior of the circle
and its harmonic conjugate. Their properties are examined analytically and through
numerical adjoint solutions. Finally, a continuous formulation of the Kutta condition for
the adjoint full-potential equations is discussed and interpreted in terms of singular
boundary forcing, Green-function kernels, and an equivalent Lagrange-multiplier
formulation.

\end{abstract}

\section{Introduction}
Potential flow methods have long been used in computational aerodynamics \cite{jameson1990}, but 
still provide a useful medium-fidelity computational fluid dynamics (CFD) modeling framework. Coupled to an integral boundary layer solver, a full-potential solver offers comparable accuracy to an Euler solver with significant reduction in computational resources. Potential flow methods can also be used in aerodynamic design applications, and in fact the very first adjoint approach described in Jameson's seminal work on adjoint methods \cite{jameson1988} was precisely a full-potential adjoint model. In this paper, we will focus on the adjoint full-potential equations. Full-potential adjoint methods have been a relevant research topic for over four decades \cite{jameson1988,angrand1983,reuther1994,reuther1996,santos2000,lewis1996,kuruvila1995,galbraith2017,crovato2023}. Applications to incompressible potential flow (i.e., flow obeying Laplace's equation) can also be found, for example, in \cite{galbraith2017,anevlavi2021,pierce2004} and, more recently, in \cite{lozano2025}. The present paper is closely related to the incompressible analysis of \cite{lozano2025}, but focuses on the compressible subcritical case, where explicit solutions are generally no
longer available and the main interest lies in the analysis of the underlying structure of the problem.

Following the derivation of the analytic lift and drag-based adjoint solutions for the 2D incompressible Euler equations in \cite{lozano2022}, a solution was proposed in \cite{lozano2023} for the lift-based compressible adjoint 2D Euler equations that contained two unknown functions encapsulating the effect on lift of perturbations to the Kutta condition. These Kutta functions are linked by a pair of equations that were written down explicitly in \cite{lozano2023} but could not be solved. In the incompressible case, on the other hand, these functions have been computed exactly and turn out to be the Poisson kernel of the Laplacian on the exterior of the circle and its harmonic conjugate \cite{lozano2025,lozano2022,krantz1999}. In this work, we will examine the relation between the 2D compressible adjoint Euler equations and the adjoint full-potential equations and use it to obtain a representation of the corresponding adjoint fields for the latter using the results in \cite{lozano2023}. This analysis will shed light on the nature of the Kutta functions and offer an interpretation for the adjoint potential and stream function, which is confirmed by examining the problem with the Green's function approach \cite{lozano2022,giles1997,giles2001}. It will also show that the equations obeyed by the Kutta functions that were obtained in \cite{lozano2023} are generalizations of the Cauchy-Riemann equations obeyed in the incompressible case, and that the Kutta functions obey the adjoint equations (which in this case are identical to the linearized full-potential equations).

A central issue in this setting is the treatment of the Kutta and wake conditions, which
the present analysis helps clarify. In potential flow, the Kutta condition is required to select the physically meaningful solution. Likewise, adjoint potential flow fundamentally requires accounting for perturbations to the Kutta/wake conditions \cite{lozano2025}. This issue has been approached from both continuous \cite{reuther1996,kuruvila1995} as well as discrete \cite{angrand1983,galbraith2017,crovato2023,anevlavi2021} perspectives, but a consistent analytic formulation is still lacking \cite{galbraith2017}. 

These insights are helpful in several directions. They contribute to understanding the mathematical foundation of adjoint equations and also provide useful benchmarks and interpretation
tools for adjoint numerical solvers. In this regard, the development and verification of
adjoint codes is a critical step in the application of adjoint methods. If exact or
partially explicit solutions are available, they provide valuable test cases; if they are
not, structural information on the governing operators, boundary conditions, and singular
behavior remains equally important. Recently, several authors have pursued both directions.
The connection between adjoint variables and Green's functions has enabled the construction
of exact adjoint solutions for quasi-one-dimensional (quasi-1D) inviscid flows \cite{giles2001,lozano2018},
two-dimensional (2D) incompressible inviscid flows \cite{lozano2025,lozano2022}, and has also helped outline
the compressible subcritical \cite{lozano2023} and transonic \cite{giles1997} inviscid cases. For inviscid 2D and 3D flows, entropy
variables provide exact solutions for computing net entropy flux across boundaries \cite{fidkowski2010,lozano2019}.
A related formulation recently developed by the authors offers an exact adjoint solution for
near-field aerodynamic drag computation \cite{lozano2021}. In the context of viscous flows, exact adjoint solutions to the Navier-Stokes equations have been proposed for both laminar \cite{kuhl2021,lozano2024} and turbulent \cite{kuhl2022} boundary layers. Two recent papers \cite{peter2022,ancourt2023} derived ordinary differential equations that adjoint solutions must satisfy along characteristic lines, enabling analytic solutions in simple cases \cite{lozano2025_2}.

The connection established here with the adjoint Euler equations makes it possible
to examine the compressible subcritical case from a different perspective. In the
incompressible setting, the relevant kernels can be obtained in closed form. In the
compressible subcritical case, by contrast, explicit formulas are generally unavailable. The
interest of the present analysis therefore lies not in a new explicit solution, but
in identifying the underlying structure of the problem: the relation between the adjoint potential and
stream function and the corresponding Euler adjoint variables, the role of the Kutta
functions within the linearized full-potential operators, and their interpretation in
terms of Green's-function kernels and enforcement of the Kutta condition in the continuous adjoint framework. It is hoped that these
results will also contribute to a more complete understanding of the continuous adjoint formulation for the full-potential equations, which is essential to determine discrete adjoint consistency. Adjoint consistent discretizations are relevant because they have optimal grid convergence properties \cite{hicken2014} and, likewise, adjoint consistency is a fundamental ingredient for adjoint-based error analysis.

To state the contribution of the present work precisely, it is useful to separate it
explicitly from the results on which it builds. The analytic adjoint solution for the 2D
incompressible Euler equations was obtained in \cite{lozano2022}, where the two Kutta
functions $\Upsilon^{(1)}$ and $\Upsilon^{(2)}$ were already identified, in the incompressible
limit, with the Poisson kernel of the Laplacian on the exterior of the circle and its harmonic
conjugate. The incompressible \emph{potential} adjoint problem was solved in closed form in
\cite{lozano2025}, where the Kutta condition was enforced through a conformal mapping of the
airfoil exterior to the exterior of a circle, and where the trailing-edge point contribution to
the adjoint solution was identified a posteriori from the closed-form solution. For the
\emph{compressible} subcritical Euler equations, \cite{lozano2023} proposed the lift-based
adjoint solution and wrote down explicitly the pair of equations satisfied by $\Upsilon^{(1)}$
and $\Upsilon^{(2)}$, but those equations were left unsolved. Separately, a variational route to
the adjoint Kutta condition in potential flow was introduced in \cite{letter}, where the
trailing-edge multiplier was stated but not derived.

The present paper concerns the compressible subcritical \emph{full-potential} adjoint equations
and contributes the following, none of which appears in the works above: (i) the exact
correspondence \eqref{eq32} between the adjoint full-potential variables and the compressible
adjoint Euler variables, valid in the elliptic sector carried by the finite wall combinations
$I^{(1)}$ and $I^{(2)}$, together with the resulting representation of the full-potential adjoint
fields obtained from \cite{lozano2023}; (ii) the identification of the Kutta-function equations of
\cite{lozano2023} as a generalized Cauchy--Riemann ($A$-conjugate) system, and the fact that the
Kutta functions themselves satisfy the linearized full-potential operators; (iii) a self-contained
derivation of the trailing-edge Kutta multiplier in the compressible full-potential setting,
carried out in both the potential and stream-function formulations, supplying the derivation that
was omitted from \cite{letter}; and (iv) the representation of the resulting singular forcing in
terms of the Poisson kernel of the linearized stream-function operator
$\mathcal{L}=\nabla\cdot(A\nabla)$ and the tangential derivative of the Neumann Green's function of
the linearized potential operator $\hat{\mathcal{L}}=\nabla\cdot(\hat A\nabla)$---the compressible,
variable-coefficient generalization of the exterior-circle Poisson kernel identified in
\cite{lozano2022}. In short, \cite{lozano2022} identified the incompressible kernels,
\cite{lozano2025} solved the incompressible potential problem and exhibited the trailing-edge term
a posteriori, \cite{lozano2023} posed but did not solve the compressible problem, and \cite{letter}
stated the variational multiplier without derivation; the present work supplies the compressible
full-potential structure, the derivation of the multiplier, and the kernel interpretation that
connect these threads.

The paper is organized as follows. Section \ref{sec:2} reviews the adjoint full-potential equations
in both the potential and stream-function formulations and discusses the generalized
Cauchy-Riemann structure relating them. Section \ref{sec:3} interprets the adjoint variables
through the Green's-function approach. Section \ref{sec:4} establishes the connection with the
compressible adjoint Euler equations. Section \ref{sec:5} discusses the Kutta functions, their
governing equations, their incompressible limit, and some heuristic aspects of their local singular structure and far-field properties.
Section \ref{sec:6} addresses the Kutta condition in the continuous adjoint setting and
interprets the Kutta functions in terms of singular boundary forcing, Green-function
kernels, and an equivalent Lagrange-multiplier formulation. Section \ref{sec:7} compares these
results with numerical adjoint solutions.

\section{The adjoint full-potential equation}
\label{sec:2}
Consider two-dimensional compressible inviscid flow. For subcritical flows, an initially irrotational flow will remain irrotational, and we can assume that the velocity vector is the gradient of a potential, $\vec{v}=\nabla\phi$. The potential flow equation is
\begin{equation}
\nabla\cdot(\rho\nabla\phi)=0 
\label{eq1}
\end{equation}
which can also be written as 
\begin{equation}
\nabla\cdot(\rho\vec{v})=\frac{\partial}{\partial x}(\rho u)+\frac{\partial}{\partial y}(\rho v)=0
\label{eq2}
\end{equation}
where $(u, v)$ represent the Cartesian velocity components and the density $\rho$ is given by
\begin{equation}
\rho=\rho_{\infty}\left(1+\frac{\gamma-1}{2}{M_{\infty}}^{2}\left(1-\frac{q^{2}}{{q_{\infty}}^{2}}\right)\right)^{\frac{1}{\gamma-1}}
\label{eq3}
\end{equation}
where $q=\sqrt{u^{2}+v^{2}}$ is the velocity magnitude, $q_{\infty}$ is the free-stream value of $q$, $M_{\infty}$ is the free-stream Mach number and $\gamma$ is the ratio of specific heats. At solid walls, the potential obeys a Neumann (zero normal derivative or non-penetration) condition $\partial_{n}\phi=0$ where $\partial_{n}$ is the normal derivative. 

In 2D, potential flow can also be described in terms of a stream function $\psi$
related to the velocity as $\rho\vec{v}=(\partial_y\psi,-\partial_x\psi)$. In the case of an
incompressible fluid with density $\rho_\infty$, $\phi$ and $\psi/\rho_\infty$ are connected
by the Cauchy-Riemann equations, so that the complex potential
$\phi+i\psi/\rho_\infty$ is an analytic function of a complex variable. Taking the real and
imaginary parts of the complex potential we obtain $\phi$ and $\psi/\rho_\infty$,
respectively. A much more complicated situation occurs in the compressible fluid case, where $\phi$ and $\psi$ are linked by the equations

\begin{align}
\frac{\partial\phi}{\partial x}&=\frac{1}{\rho}\frac{\partial\psi}{\partial y} \nonumber \\
\frac{\partial\phi}{\partial y}&=-\frac{1}{\rho}\frac{\partial\psi}{\partial x}
\label{eq4}
\end{align}
which generalize the Cauchy-Riemann equations to the compressible case. In terms of the stream function, the flow equation is
\begin{equation}
\nabla\cdot({\rho}^{-1}\nabla\psi)=0
\label{eq5}
\end{equation}
which is simply the irrotationality condition $\nabla\times\vec{v}=\partial_{x}v-\partial_{y}u=0$. At solid walls, the stream function obeys a Dirichlet (prescribed value) condition $\psi=$ constant. 

We will focus, for definiteness, on steady, two-dimensional, subcritical potential flow on a domain $\Omega$ with far-field boundary $S^{\infty}$ and (closed) wall boundary $S$ (typically an airfoil profile). We will be considering the adjoint problem for an objective function measuring the non-dimensional aerodynamic force on $S$ along a direction $\vec{d}$, 
\begin{equation}
I=\int_{S}c_{\infty}^{-1}(p-p_{\infty})(\hat{n}\cdot\vec{d})ds=\int_{S}c_{\infty}^{-1}p(\hat{n}\cdot\vec{d})ds
\label{eq6}
\end{equation}
where $p$ is the pressure, $\hat{n}$ is the unit normal to $S$ pointing towards the interior of $S$ and $c_{\infty}=\rho_{\infty} q_{\infty}{}^{2}l/2$ is a normalization constant, with $l$ a reference length (equal to the chord length for an airfoil or to the diameter for a circle). 

The adjoint problem is constructed by augmenting eq. (\ref{eq6}) with the flow equations weighted by a Lagrange multiplier (the adjoint variable). In the present case, there are two different versions of the flow equations, so the adjoint problem can be defined in either of two ways by considering the potential or the stream function.

\subsection{Adjoint potential}
\label{subsec:21}
Using the velocity potential, the flow equation is (\ref{eq1}) with boundary conditions $\partial_{n}\phi_{s}=0$ at the wall and $\phi\rightarrow q_{\infty}\vec{x}\cdot(\cos~\alpha,\sin~\alpha)$ at the far field, where $\alpha$ (the angle of attack) sets the flow direction far from the body. Using a Lagrange multiplier $\tilde{\phi}$ to enforce the flow equation yields the following Lagrangian
\begin{equation}
L=\int_{S}c_{\infty}^{-1}p(\hat{n}\cdot\vec{d})ds-\int_{\Omega}\tilde{\phi}\nabla\cdot(\rho\vec{v})d\Omega=\int_{S}c_{\infty}^{-1}p(\hat{n}\cdot\vec{d})ds-\int_{\Omega}\tilde{\phi}\nabla\cdot(\rho\nabla\phi)d\Omega
\label{eq7}
\end{equation}

The Lagrange multiplier $\tilde{\phi}$ is the adjoint potential. In order to derive the adjoint equations and boundary conditions, it suffices to linearize (\ref{eq7}) with respect to flow variations $\delta\phi$, thus ignoring the effect of shape variations, which only contribute to the sensitivities and not to the adjoint problem formulation. We will also assume that the circulation of the flow around the airfoil is fixed, thus leaving out, for the time being, the contribution of the Kutta condition and the wake cut. 

To linearize (\ref{eq7}) with respect to flow variations, we will need the linearization of $\rho$ and $p$ in terms of $\delta\phi$. For isentropic flow, $\delta p=a^{2}\delta\rho$, where $a$ is the sound speed, and using (\ref{eq3}) $\delta\rho$ is given by
\begin{equation}
\delta\rho=-\rho a^{-2}\nabla\phi\cdot\nabla\delta\phi
\label{eq8}
\end{equation}

Using (\ref{eq8}) to linearize the potential equation (\ref{eq1}) yields 
\begin{equation}
0=\nabla\cdot(\delta\rho\nabla\phi+\rho\nabla\delta\phi)=\partial_{i}(-\rho a^{-2}\partial_{i}\phi\partial_{j}\phi\partial_{j}\delta\phi+\rho\partial_{i}\delta\phi)=\partial_{i}(\hat{A}_{ij}\partial_{j}\delta\phi)
\label{eq9}
\end{equation}
where
\begin{equation}
\hat{A}_{ij}=\rho(\delta_{ij}-u_{i}u_{j}/a^{2})
\label{eqAhat}
\end{equation}
Here, $\delta_{ij}$ is the Kronecker delta and indices $i, j$ are used to denote the Cartesian coordinates and velocity components and we follow the convention that summation over $i, j=1$ to $2$ is implied by a repeated index. Eq. (\ref{eq9}) is the potential equation linearized with respect to small perturbations $\delta\phi$ about a general flow.  It is a self-adjoint elliptic equation for subsonic flow conditions and it is different from the more conventional small-disturbance potential equation linearized around a uniform flow. 

Using (\ref{eq9}) and the flow boundary condition $\partial_{n}\phi_{s}=0$ to linearize (\ref{eq7}) and integrating by parts in both the adjoint-weighted linearized equation and the cost function yields
\begin{align}
\delta L &=-\int_{S}c_{\infty}^{-1}\rho\nabla\phi\cdot\nabla\delta\phi(\hat{n}\cdot\vec{d})ds-\int_{\Omega}\tilde{\phi}\partial_{i}(\hat{A}_{ij}\partial_{j}\delta\phi)d\Omega \nonumber \\
&=\int_{S}c_{\infty}^{-1}\partial_{s}(\rho(\hat{n}\cdot\vec{d})\partial_{s}\phi)\delta\phi ds-\int_{\Omega}\partial_{j}(\hat{A}_{ij}\partial_{i}\tilde{\phi})\delta\phi d\Omega \nonumber \\
&\quad +\int_{S}\rho\partial_{n}\tilde{\phi}\delta\phi ds+\int_{S_{\infty}}\partial_{i}\tilde{\phi}\hat{A}_{ij}n_{j}\delta\phi ds-\int_{S}\rho\tilde{\phi}\partial_{n}\delta\phi ds \nonumber \\
&\quad -\int_{S_{\infty}}\tilde{\phi}n_{i}\hat{A}_{ij}\partial_{j}\delta\phi ds
\label{eq10}
\end{align}

The adjoint equation and boundary conditions are obtained from (\ref{eq10}) by eliminating the coefficients of the unconstrained flow variations $\delta\phi$. The volume term is canceled if $\tilde\phi$ satisfies the adjoint equation 
\begin{equation}
\partial_{j}(\hat{A}_{ij}\partial_{i}\tilde{\phi})=0 \quad {\rm in}~\Omega
\label{eq11}
\end{equation}
(which is identical to the linearized potential equation obeyed by $\delta\phi$). At the far field, the perturbation potential \(\delta\phi\) is prescribed, whereas the conormal derivative \(n_{i}\hat{A}_{ij}\partial_{j}\delta\phi\) is unconstrained; its coefficient is therefore eliminated by imposing \(\tilde\phi\to 0\). On the wall, the non-penetration condition fixes \(n_{i}\hat{A}_{ij}\partial_{j}\delta\phi\equiv \rho\partial_n\delta\phi\), whereas the wall value of \(\delta\phi\) is arbitrary. Canceling its coefficient gives the adjoint wall boundary condition
\begin{equation}
\partial_{n}\tilde{\phi}=-\rho^{-1}\partial_{s}(\rho(\hat{n}\cdot\vec{d})\partial_{s}\phi)/c_{\infty}
\label{eq12}
\end{equation}
Finally, terms involving prescribed boundary variations, such as $\partial_n\delta\phi$ on the wall, are retained and give the adjoint contribution to the sensitivity once the adjoint problem has been solved.

In (\ref{eq10}), we have overlooked the endpoint contribution
\begin{equation}
-\left[c_\infty^{-1}\rho\partial_{s}\phi(\hat{n}\cdot\vec{d})\delta\phi\right]_{s=0}^{s_{\rm max}}
\label{eq13}
\end{equation}
that results from the integration by parts of the linearized cost function along the wall boundary. Since at this stage circulation variations are excluded, $\delta\phi$ is single-valued on the closed wall contour and the above contribution vanishes. The case in which circulation is allowed to vary is treated separately in section \ref{sec:6}.



\subsection{Adjoint stream function}
Using instead the stream-function formulation (\ref{eq5}), the Lagrangian takes the form
\begin{equation}
L=\int_{S}c_{\infty}^{-1}p(\hat{n}\cdot\vec{d})ds+\int_{\Omega}\tilde{\psi}(\nabla\times\vec{v})d\Omega=\int_{S}c_{\infty}^{-1}p(\hat{n}\cdot\vec{d})ds-\int_{\Omega}\tilde{\psi}\nabla\cdot({\rho}^{-1}\nabla\psi)d\Omega
\label{eq15}
\end{equation}
where $\nabla\times\vec{v}=\partial_{x}v-\partial_{y}u$ and $\tilde{\psi}$ is the adjoint stream function. 

Linearization of eq. (\ref{eq15})  with respect to $\delta\psi$ requires $\delta\rho$, which in terms of the stream function is given 
\begin{equation}
\delta\rho=-\frac{\nabla\psi\cdot\nabla\delta\psi}{\rho a^{2}(1-M^{2})}
\label{eq16}
\end{equation}
We can use (\ref{eq16}) to obtain the linearized flow equation
\begin{equation}
0=\delta\nabla\cdot({\rho}^{-1}\nabla\psi)=\partial_{i}\left(A_{ij}\partial_{j}\delta\psi\right)
\label{eq17}
\end{equation}
where
\begin{equation}
A_{ij}={\rho}^{-1}\delta_{ij}+\frac{\partial_{i}\psi\partial_{j}\psi}{{\rho}^{3}a^{2}(1-M^{2})}
\label{eq59}
\end{equation}

Linearizing (\ref{eq15}) with respect to flow variations using (\ref{eq16}) and (\ref{eq17}) and integrating by parts yields
\begin{align}
\delta L&=-\int_{S}c_{\infty}^{-1}
\frac{\nabla\psi\cdot\nabla\delta\psi}{\rho(1-M^{2})}
(\hat{n}\cdot\vec{d})ds
-\int_{\Omega}\tilde{\psi}\partial_i(A_{ij}\partial_j\delta\psi)d\Omega
\nonumber \\
&=-\int_{S}c_{\infty}^{-1}
\frac{\partial_n\psi\,\partial_n\delta\psi}{\rho(1-M^{2})}
(\hat{n}\cdot\vec{d})ds
-\int_{\Omega}\delta\psi\,\partial_j(A_{ij}\partial_i\tilde{\psi})d\Omega
\nonumber \\
&\quad
-\int_S \rho^{-1}(1-M^2)^{-1}\tilde\psi\,\partial_n\delta\psi\,ds
-\int_{S_\infty}\tilde\psi\,n_iA_{ij}\partial_j\delta\psi\,ds
\nonumber \\
&\quad
+\int_S \rho^{-1}(1-M^2)^{-1}\partial_n\tilde\psi\,\delta\psi\,ds
+\int_{S_\infty}\delta\psi\,n_jA_{ij}\partial_i\tilde\psi\,ds .
\label{eq18}
\end{align}

As in the potential case, we can deduce from (\ref{eq18}) that the adjoint stream function obeys the equation 
\begin{equation}
\partial_{j}\left(A_{ij}\partial_{i}\tilde{\psi}\right)=0 \quad \text{in } \Omega ,
\label{eq19}
\end{equation}
a suitable vanishing condition at infinity (involving $\tilde\psi$ or its conormal derivative depending on whether $\delta\psi$ or its conormal derivative is prescribed), and the wall boundary condition 
\begin{equation}
\tilde{\psi} = -c_\infty^{-1}\partial_n\psi(\hat n\cdot\vec d) \quad \text{on } S,
\label{eq20}
\end{equation}
where the identity $n_{i}A_{ij}\partial_{j}\delta\psi={\rho}^{-1}(1-M^{2})^{-1}\partial_{n}\delta\psi$ at the wall has been used.

\subsection{Relation between the potential and stream-function formulations}
\label{ssec:conjugacy}

We have mentioned above that the relation between the adjoint potential and stream-function formulations generalizes the relation between the real and imaginary parts of a holomorphic function. Let us now make these claims precise.   

\subsubsection{The generalized Cauchy-Riemann equations} 

Let us define the operator ${\cal L} = \nabla \cdot (A\nabla)$ where $A$ is given by (\ref{eq59}). For the Laplace equation, the harmonic conjugate $v$ of a harmonic function $u$ satisfies the Cauchy-Riemann equations: $\nabla u = J \nabla v$, where $J = \begin{pmatrix} 0 & 1 \\ -1 & 0 \end{pmatrix}$ is the 90-degree rotation matrix. For the generalized operator ${\cal L}$, we say that $(\tilde\phi,\tilde\psi)$ form an $A$-conjugate pair \cite{Astala2009} if they obey the generalized Cauchy-Riemann equations
\begin{equation}
A\nabla \tilde\psi = J\nabla \tilde\phi ,
\label{eq:GenCR}
\end{equation}
Equation \eqref{eq:GenCR} reduces to the classical Cauchy-Riemann equations when $A=I$.


We can now prove that if $(\tilde\phi,\tilde\psi)$ is an $A$-conjugate pair obeying \eqref{eq:GenCR}, then $\tilde\psi$ obeys the adjoint stream-function equation (\ref{eq19}) while $\tilde\phi$ obeys the adjoint potential equation (\ref{eq11}). The first assertion follows by taking the divergence of \eqref{eq:GenCR}, which immediately yields
\begin{equation}
\nabla\cdot(A\nabla\tilde\psi)=\nabla\cdot(J\nabla\tilde\phi)=0 .
\end{equation}
On the other hand, solving \eqref{eq:GenCR} for $\nabla\tilde\psi$ gives $\nabla\tilde\psi=A^{-1}J\nabla\tilde\phi$. 
Multiplying by $J$ and taking the divergence, we obtain the following equation for $\tilde\phi$
\begin{equation}
0=\nabla\cdot(J\nabla\tilde\psi)
   =\nabla\cdot(JA^{-1}J\nabla\tilde\phi)
\end{equation}
Since $A$ is a $2\times 2$ symmetric matrix, $JA^{-1}J= -A/\det A$ and therefore $\tilde\phi$ satisfies the conjugate equation
\begin{equation}
\nabla\cdot\!\left(\frac{A}{\det A}\nabla\tilde\phi\right)=0
\label{eq:conjEq1}
\end{equation}
 
Using the identity $\hat{A} = A/\det A$, (\ref{eq:conjEq1}) becomes
\begin{equation}
\hat{{\cal L}} \tilde\phi = \nabla \cdot (\hat{A} \nabla \tilde\phi) = 0 
\label{eq:conjEq2}
\end{equation}
This shows that the adjoint potential and stream function are related as conjugate variables for the pair of elliptic operators $\hat{{\cal L}}$ and $\cal{L}$. Hence the generalized Cauchy-Riemann system \eqref{eq:GenCR} couples solutions of the
adjoint stream-function and adjoint potential equations.

\section{Green's function approach}
\label{sec:3}
The adjoint solution can be interpreted as the influence of a point-source on the objective function. If this influence can be independently computed, it can be used to obtain exact predictions for the adjoint variables. This is the essence of Giles and Pierce \cite{giles1997,giles2001} Green's function approach, which was used in \cite{lozano2022} and \cite{lozano2025} to compute the analytic adjoint solution for the 2D incompressible Euler equations and 2D incompressible potential flow, respectively. For compressible potential flow, there are two independent adjoint states and, correspondingly, two Green's functions that can be considered. These were identified in \cite{giles1997} as a mass source injecting fluid with the local values of stagnation pressure and enthalpy, which is the analog of the incompressible potential point source, and as an applied force in the direction normal to the local flow, which is the analog of the incompressible point vortex. The linearized force exerted by these Green's functions is derived in Appendix \ref{app:A}. 

\subsection{Green's function for the adjoint potential}
We begin with the potential formulation given by eq. (\ref{eq7}). Adding a point source perturbation to the linearized continuity equation we get the Lagrangian 
\begin{equation}
\delta L=\int_{S}c_{\infty}^{-1}\delta p(\hat{n}\cdot\vec{d})ds-\int_{\Omega}\tilde{\phi}(\nabla\cdot\delta(\rho\vec{v})-\delta(\vec{x}-\vec{x}_{0}))d\Omega
\label{eq21}
\end{equation} 
where $\delta(\vec{x}-\vec{x}_{0})$ is the Dirac delta function. If $\delta\vec{v}$ is subject to homogeneous boundary conditions and if $\tilde{\phi}$ obeys the adjoint equation (\ref{eq11}) and the wall boundary condition (\ref{eq12}), eq. (\ref{eq21}) yields 
\begin{equation}
\delta L=\tilde{\phi}(\vec{x}_{0})
\label{eq22}
\end{equation} 
which can be used to obtain the value of $\tilde{\phi}(\vec{x}_{0})$ from $\delta L$, which is the additional force exerted by the fluid on the body due to the presence of the point perturbation. In order to compute $\delta L$, we start by noticing that the perturbation obeys the equation 
\begin{equation}
\nabla\cdot\delta(\rho\vec{v})=\delta(\vec{x}-\vec{x}_{0})
\label{eq23}
\end{equation}
which describes a linearized compressible point mass source. Hence, the adjoint potential at a point represents the linearized force due to a compressible mass source inserted at that point. This makes sense given that in section \ref{sec:2} we introduced the adjoint potential as the dual variable to the continuity equation. No closed-form solution of eq. (\ref{eq23}) is available in general. However, the force exerted by the solution of eq. (\ref{eq23}) was obtained in \cite{lozano2023} and is given by eq. (\ref{eq81}) and (\ref{eq83}) below, yielding the following result for the drag and lift-based adjoint potential solutions 
\begin{align}
\tilde{\phi}_{D}&=\frac{1}{c_{\infty}}(u\cos\alpha+v\sin\alpha-q_{\infty}) \nonumber \\
\tilde{\phi}_{L}&=\frac{1}{c_{\infty}}(v\cos\alpha-u\sin\alpha)+\frac{q_{\infty}}{c_{\infty}}\Upsilon^{(1)}
\label{eq24}
\end{align} 
The drag solution is completely determined and is smooth throughout the flow field, being essentially the projection of the momentum along the direction of the free stream. The lift solution, on the other hand, contains an a priori undetermined function $\Upsilon^{(1)}$, which physically represents the amount of lift required to restore the Kutta condition in the presence of a mass source, which will be explored in more depth in section \ref{sec:5}.

\subsection{Green's function for the adjoint stream function}
We next turn to the stream-function formulation. Adding a point source perturbation to the linearized irrotationality equation modifies the Lagrangian (\ref{eq15}) as follows
\begin{equation}
\delta L=\int_{S}c_{\infty}^{-1}\delta p(\hat{n}\cdot\vec{d})ds+\int_{\Omega}\tilde{\psi}(\nabla\times\delta\vec{v}+\delta(\vec{x}-\vec{x}_{0}))d\Omega
\label{eq25}
\end{equation}
If the boundary conditions for $\delta\vec{v}$ are again homogeneous, eq. (\ref{eq25}) yields, after integrating by parts and using the adjoint equation (\ref{eq19}) and boundary condition (\ref{eq20}), the following result
\begin{equation}
\delta L=\tilde{\psi}(\vec{x}_{0})
\label{eq26}
\end{equation}
which can be used to obtain the value of $\tilde{\psi}(\vec{x}_{0})$. The perturbation obeys the equation
\begin{equation}
\nabla\times\delta\vec{v}=\partial_{x}\delta v-\partial_{y}\delta u=-\delta(\vec{x}-\vec{x}_{0})
\label{eq27}
\end{equation}
which represents a (compressible) point vortex. Hence, the adjoint stream function at a point (which is dual to the vorticity equation) represents the force due to a compressible vortex inserted at that point. No closed-form solution of eq. (\ref{eq27}) is available in general, but we can use eq. (\ref{eq81}) and (\ref{eq83}) for $\delta L$ in eq. (\ref{eq26}) and write down the analytic stream-function solutions as 
\begin{align}
\tilde{\psi}_{D}&=\frac{\rho}{c_{\infty}}(v\cos\alpha-u\sin\alpha) \nonumber \\
\tilde{\psi}_{L}&=-\frac{\rho}{c_{\infty}}(u\cos\alpha+v\sin\alpha)+\frac{\rho q_{\infty}}{c_{\infty}}(1+\Upsilon^{(2)})
\label{eq28}
\end{align}
Again, the drag-based solution is smooth throughout the flow field (it consists essentially of the projection of the momentum along the direction perpendicular to the free stream), while the lift-based solution contains a second unknown function $\Upsilon^{(2)}$ that represents the response of the flow to the perturbation to the Kutta condition due to the point source (\ref{eq27}) (see section \ref{sec:5} for further details).

\section{Relation to the 2D compressible adjoint Euler equations}
\label{sec:4}
For subcritical irrotational flows, solutions to the full-potential equations are also solutions to the Euler equations. It is interesting to ask for the possible relation between the corresponding adjoint states, which would allow to interpret the adjoint potential/stream function in terms of the more conventional 4-component adjoint state of the Euler equations and to obtain a representation for the adjoint potential solutions from the solution to the compressible adjoint Euler equations obtained in \cite{lozano2023}. Likewise, in section \ref{sec:7} we will use this relation to obtain numerical solutions for the adjoint potential equations from numerical adjoint solutions to the compressible Euler equations. 

For completeness, we start by recalling the definition of the steady adjoint Euler equations. The primal flow is governed by the compressible Euler equations $R(U)=\nabla\cdot F(U)=\partial_{x}F_{x}+\partial_{y}F_{y}=0$, where
\begin{equation}
U=\begin{pmatrix}\rho\\ \rho u\\ \rho v\\ \rho E\end{pmatrix}, \quad F_{x}=\begin{pmatrix}\rho u\\ \rho u^{2}+p\\ \rho uv\\ \rho uH\end{pmatrix}, \quad F_{y}=\begin{pmatrix}\rho v\\ \rho uv\\ \rho v^{2}+p\\ \rho vH\end{pmatrix}
\label{eq29}
\end{equation}
and $E$ and $H$ are the total energy and enthalpy, respectively. To obtain the sensitivities of the cost function (\ref{eq6}) with respect to flow perturbations $\delta U$ using the adjoint approach we augment the objective function with the flow equations with a vector of Lagrange multipliers $\psi=(\psi_{\rho},\psi_{x},\psi_{y},\psi_{E})^{T}$ (the adjoint variables), which yields the Lagrangian


\begin{align}
L&=\int_{S}c_{\infty}^{-1}(p-p_{\infty})(\hat{n}\cdot\vec{d})ds-\int_{\Omega}\psi^{T}\nabla\cdot\begin{pmatrix}\rho\vec{v}\\ \rho u\vec{v}+\vec{v}p\\ \rho v\vec{v}+\vec{v}p\\ \rho H\vec{v}\end{pmatrix}d\Omega \nonumber \\
&=\int_{S}c_{\infty}^{-1}(p-p_{\infty})(\hat{n}\cdot\vec{d})ds-\int_{\Omega}\psi^{T}\begin{pmatrix}1\\ u\\ v\\ H\end{pmatrix}\nabla\cdot(\rho\vec{v})d\Omega-\int_{\Omega}\psi^{T}\begin{pmatrix}0\\ \rho\vec{v}\cdot\nabla u+\partial_{x}p\\ \rho\vec{v}\cdot\nabla v+\partial_{y}p\\ \rho\vec{v}\cdot\nabla H\end{pmatrix}d\Omega
\label{eq30}
\end{align}
where we have separated $\nabla\cdot\vec{F}$ in two terms for future convenience. The non-linear flow equations ensure that $\vec{v}\cdot\nabla H=0$. Likewise, for homentropic flow $\partial_{i}p=-\rho\vec{v}\cdot\partial_{i}\vec{v}$, so eq. (\ref{eq30}) can be written as 
\begin{align}
L&=\int_{S}c_{\infty}^{-1}(p-p_{\infty})(\hat{n}\cdot\vec{d})ds-\int_{\Omega}\psi^{T}\begin{pmatrix}1\\ u\\ v\\ H\end{pmatrix}\nabla\cdot(\rho\vec{v})d\Omega-\int_{\Omega}\psi^{T}\begin{pmatrix}0\\ -\rho v\\ \rho u\\ 0\end{pmatrix}(\nabla\times\vec{v})d\Omega \nonumber \\
&=\int_{S}c_{\infty}^{-1}(p-p_{\infty})(\hat{n}\cdot\vec{d})ds-\int_{\Omega}(\psi_{\rho}+u\psi_{x}+v\psi_{y}+H\psi_{E})\nabla\cdot(\rho\vec{v})d\Omega \nonumber \\
&\quad +\int_{\Omega}\rho(v\psi_{x}-u\psi_{y})(\nabla\times\vec{v})d\Omega
\label{eq31}
\end{align}
Comparing eq. (\ref{eq31}) with eq. (\ref{eq7}) and eq. (\ref{eq15}) we find the following correspondence between the potential and Euler compressible adjoint states 
\begin{align}
\tilde{\phi}&=\psi_{\rho}+u\psi_{x}+v\psi_{y}+H\psi_{E} \nonumber \\
\tilde{\psi}&=\rho(v\psi_{x}-u\psi_{y})
\label{eq32}
\end{align} 
Eq. (\ref{eq32}) relates the adjoint potential with $I^{(1)}=\psi_{\rho}+u\psi_{x}+v\psi_{y}+H\psi_{E}$ and the adjoint stream function with $I^{(2)}=-\rho(v\psi_{x}-u\psi_{y})$. $I^{(1)}$ and $I^{(2)}$ were identified in \cite{giles1997} as the value of the linearized cost function due to a point mass source and a point force normal to the local flow direction, respectively. Eq. (\ref{eq32}) thus confirms that the adjoint potential variables are related to the mass and force Green-function perturbations discussed in section \ref{sec:3}. 

$I^{(1)}$ and $I^{(2)}$ are also related to the continuous adjoint expression for the sensitivity derivatives and to the adjoint wall boundary condition, respectively, which motivates a second comment. As explained in \cite{lozano2022,lozano2023,lozano2019_2}, lift-based adjoint solutions around airfoils diverge not only at the rear stagnation point/trailing edge, but also along the incoming stagnation streamline and the wall. There are, however, two combinations of adjoint variables that have a finite value along the wall. These are, precisely, $I^{(1)}$ and $I^{(2)}$, which, as we have just seen, correspond to the solutions to the adjoint full-potential equation. The origin of the streamline singularity is in fact in a non-potential perturbation $I^{(4)}$ that changes the total pressure of the flow.  

Finally, eq. (\ref{eq32}) can also be used to derive the adjoint potential and stream function from the  solutions for $(\psi_{\rho},\psi_{x},\psi_{y},\psi_{E})$ obtained in \cite{lozano2021,lozano2023}. These are 
\begin{equation}
\Psi_{D}=\begin{pmatrix}
\psi_{\rho} \\
\psi_{x} \\
\psi_{y} \\
\psi_{E}
\end{pmatrix}_{\!\! D}=\frac{1}{\rho_{\infty}^{1-\gamma}q_{\infty}c_{\infty}}\begin{pmatrix}H\rho^{1-\gamma}-H_{\infty}\rho_{\infty}^{1-\gamma}\\ -\rho^{1-\gamma}u+\rho_{\infty}^{1-\gamma}q_{\infty}\cos \alpha\\ -\rho^{1-\gamma}v+\rho_{\infty}^{1-\gamma}q_{\infty}\sin \alpha\\ \rho^{1-\gamma}-\rho_{\infty}^{1-\gamma}\end{pmatrix}
\label{eq80}
\end{equation}
for drag, where $\gamma$ is the ratio of specific heats, and 
\begin{equation}
\Psi_{L}=\begin{pmatrix}
\psi_{\rho} \\
\psi_{x} \\
\psi_{y} \\
\psi_{E}
\end{pmatrix}_{\!\! L}
= \frac{q_{\infty}}{c_{\infty}}
\begin{pmatrix}
\frac{(\gamma-1)\rho}{2\gamma p}(2\Upsilon^{(1)} - q^{2}\Xi)H \\[12pt]
-\frac{\sin\alpha}{q_{\infty}} + \frac{(\gamma-1)\rho H}{\gamma p}u\Xi - \frac{(\gamma-1)\rho}{\gamma p}(H+\frac{1}{2}q^{2})\frac{u}{q^{2}}\Upsilon^{(1)} + \frac{v}{q^{2}}(1+\Upsilon^{(2)}) \\[12pt]
\frac{\cos\alpha}{q_{\infty}} + \frac{(\gamma-1)\rho H}{\gamma p}v\Xi - \frac{(\gamma-1)\rho}{\gamma p}(H+\frac{1}{2}q^{2})\frac{v}{q^{2}}\Upsilon^{(1)} - \frac{u}{q^{2}}(1+\Upsilon^{(2)}) \\[12pt]
\frac{(\gamma-1)\rho}{2\gamma p}(2\Upsilon^{(1)} - q^{2}\Xi)
\end{pmatrix}
\label{eq33}
\end{equation}
for lift, where $\Xi$ is an integration over the local streamline of secondary mass and vorticity sources (see \cite{giles1997} \cite{lozano2022} \cite{lozano2023}). Introducing eq. (\ref{eq80}) and (\ref{eq33}) into (\ref{eq32}) we obtain 
\begin{align}
\tilde{\phi}_{D}&=\frac{1}{c_{\infty}}(u\cos\alpha+v\sin\alpha-q_{\infty}) \nonumber \\
\tilde{\psi}_{D}&=\frac{\rho}{c_{\infty}}(v\cos\alpha-u\sin\alpha)
\label{eq34}
\end{align}
for drag, and
\begin{align}
\tilde{\phi}_{L}&=\frac{1}{c_{\infty}}(v\cos\alpha-u\sin\alpha+q_{\infty}\Upsilon^{(1)}) \nonumber \\
\tilde{\psi}_{L}&=-\frac{\rho}{c_{\infty}}(u\cos\alpha+v\sin\alpha-q_{\infty}(1+\Upsilon^{(2)}))
\label{eq35}
\end{align}
for lift, which agree with (\ref{eq24}) and (\ref{eq28}).

\section{Remarks on the Kutta functions}
\label{sec:5}
The above analytic solutions are written in terms of the functions $\Upsilon^{(1)}$ and $\Upsilon^{(2)}$. These are two unknown functions related to perturbations to the Kutta condition that in the incompressible case turn out to be the Poisson kernel for the circle and its conjugate \cite{lozano2025}. In the compressible case, it was established in \cite{lozano2023} that $\Upsilon^{(1)}$ and $\Upsilon^{(2)}$ must obey the following equations
\begin{align}
(M^{2}-1)\rho\vec{v}\cdot\nabla\Upsilon^{(1)}+(v,-u)\cdot\nabla(\rho(1+\Upsilon^{(2)}))&=0 \nonumber \\
\rho(v,-u)\cdot\nabla\Upsilon^{(1)}+\vec{v}\cdot\nabla(\rho(1+\Upsilon^{(2)}))&=0
\label{eq36}
\end{align}
in $\Omega$ and $\Upsilon^{(2)}=-1$ on walls, and both functions must approach zero at infinity. Likewise, it was argued in \cite{lozano2023} that these functions must be singular at rear stagnation points or trailing edges in order for the compressible lift-adjoint state to exhibit the singularities that are observed in numerical computations and that have a direct counterpart in the incompressible case for which analytic adjoint solutions are known. Eq. (\ref{eq36}) can be written in streamline coordinates as 
\begin{align}
(M^{2}-1)\rho\partial_{s}\Upsilon^{(1)}+\partial_{n}(\rho(1+\Upsilon^{(2)}))&=0 \nonumber \\
\rho\partial_{n}\Upsilon^{(1)}+\partial_{s}(\rho(1+\Upsilon^{(2)}))&=0
\label{eq37}
\end{align}
where $s$ is the coordinate along streamlines and $n$ is the coordinate perpendicular to streamlines. Notice that, since $\Upsilon^{(2)}=-1$ at the wall, it follows from the second equation that $\partial_{n}\Upsilon^{(1)}=0$ at walls. 

In the incompressible case, $\Upsilon^{(1)}$ and $\Upsilon^{(2)}$ are the imaginary and real parts of a meromorphic function and, thus, obey the Laplace's and Cauchy-Riemann equations $\partial_{x}\Upsilon^{(1)}=-\partial_{y}\Upsilon^{(2)}$ and $\partial_{y}\Upsilon^{(1)}=\partial_{x}\Upsilon^{(2)}$. Multiplying alternatively $\nabla\Upsilon^{(1)}$ and $\nabla\Upsilon^{(2)}$ by the velocity vector $\vec{v}$ and using the Cauchy-Riemann equations leads to Equation (\ref{eq36}) with constant $\rho$ and $M=0$. 
With the trailing edge mapped to the rear stagnation point $\zeta_{te}$ on a circle centered at $\zeta_0$, where $\zeta$ is a complex coordinate parameterizing the exterior of the circle, $\Upsilon^{(2)}$ (or rather 1+$\Upsilon^{(2)}$) is the Poisson kernel for the exterior Dirichlet
problem and $\Upsilon^{(1)}$ its harmonic conjugate \cite{lozano2025,lozano2022},
\begin{equation}
\Upsilon^{(1)}_{\mathrm{inc}}(\zeta)
 =-i\left(\frac{\zeta_{te}-\zeta_0}{\zeta-\zeta_{te}}-c.c.\right)
,
\qquad
\Upsilon^{(2)}_{\mathrm{inc}}(\zeta)
 = \frac{\zeta_{te}-\zeta_0}{\zeta-\zeta_{te}}+c.c.,
\label{eq:incUpsilon}
\end{equation}
where $c.c.$ stands for complex conjugate.

In the compressible case, $\rho(1+\Upsilon^{(2)})$ and $\Upsilon^{(1)}$ form an
$A$-conjugate pair, up to the sign convention in \eqref{conjugate}, in the sense of
Section~\ref{ssec:conjugacy}. In particular, they satisfy the generalized Laplace
equations 
\begin{align}
\nabla\cdot\left(\hat{A}\nabla\Upsilon^{(1)}\right)&=0 \nonumber \\
\nabla\cdot\left(A\nabla\left(\rho\left(1+\Upsilon^{(2)}\right)\right)\right)&=0
\label{eq38}
\end{align}
and the generalized Cauchy-Riemann equations  
\begin{equation}
\hat{A}\nabla\Upsilon^{(1)}=\rho\left(1-M^2\right)\nabla\Upsilon^{(1)}+\frac{\rho}{a^2}\vec{v}^{\perp}\left(\vec{v}^{\perp}\cdot\nabla\Upsilon^{(1)}\right)=-J\nabla\left(\rho\left(1+\Upsilon^{(2)}\right)\right)
\label{conjugate}
\end{equation}
where $\vec{v}^{\perp}=J\vec{v}$, and
\begin{equation}
A\nabla\left(\rho\left(1+\Upsilon^{(2)}\right)\right)= J\nabla\Upsilon^{(1)} .
\label{conjugate2}
\end{equation}
Projecting either (\ref{conjugate}) or (\ref{conjugate2}) onto $\vec{v}$ and $\vec{v}^{\perp}$ yields eq. (\ref{eq36}).

We shall also see in section \ref{ssec:kernel} that the Kutta functions can be interpreted as kernels for the generalized Laplacians eq. (\ref{eq38}). $\rho(1+\Upsilon^{(2)})$ is the Poisson kernel for ${\cal L}=\nabla\cdot(A\nabla)$, while $\Upsilon^{(1)}$ is related to the derivative of the Neumann Green's function for $\hat{\cal L}=\nabla\cdot(\hat{A}\nabla)$. 

Getting further analytic results for $\Upsilon^{(1)}$ and $\Upsilon^{(2)}$ seems to be out of reach. A frozen-coefficient analysis suggests that the local singular structure should remain
closely related to the incompressible one. At a rear stagnation point, the frozen operator
reduces to its incompressible form, so the leading singular behavior is expected to agree with that of the incompressible kernels, with compressibility entering only through higher-order
corrections. At a cusped trailing edge, on the other hand, the frozen operator has a
Prandtl-Glauert-type form, and the local kernels are therefore expected to differ from the
incompressible ones by an anisotropic rescaling in the streamline direction by a factor
$1/\beta_{TE}$, with $\beta_{TE}=\sqrt{1-M_{TE}^2}$. 

Likewise, the far-field behavior is expected to be governed by the operators frozen at
their free-stream values. In particular, the compressible scalar kernels should decay like
$O(1/r)$ as $r\to\infty$, as in the incompressible case, but with level sets that are
elliptically distorted relative to their incompressible counterparts, the distortion being
set by the free-stream Prandtl-Glauert factor
$\beta_\infty=\sqrt{1-M_\infty^2}$. 

Additionally, for weakly subsonic flows, we can try a recursive approach \cite{vandyke1975} or a Janzen-Rayleigh expansion \cite{jacob1959} in powers of the free-stream Mach number $M_{\infty}$. The idea is to expand every field in even powers of the free stream Mach number, substitute the expansions into eqs. (\ref{eq38}) and isolate the different powers of the Mach number. This leads to a recursive set of Poisson equations at each order, the source term being a function of the lower-order solutions and their derivatives. At zeroth order, one obtains Laplace’s equations for $\Upsilon_{0}^{(i)}$ 
\begin{align}
\nabla^{2}\Upsilon_{0}^{(1)}&=0 \nonumber \\
\nabla^{2}\Upsilon_{0}^{(2)}&=0
\label{eq39}
\end{align}
corresponding to the incompressible limit $M_{\infty}\rightarrow 0$. To order ${M_{\infty}}^{2}$ we get two Poisson equations for $\Upsilon_{1}^{(i)}$ 
\begin{align}
\nabla^{2}\Upsilon_{1}^{(1)}&=-\nabla\rho_{1}\cdot\nabla\Upsilon_{0}^{(1)}/\rho_{\infty}+\partial_{j}(v_{0i}v_{0j}\partial_{i}\Upsilon_{0}^{(1)})/{q_{\infty}}^{2} \nonumber \\
\nabla^{2}\Upsilon_{1}^{(2)}&=-(1+\Upsilon_{0}^{(2)})\nabla^{2}\rho_{1}/\rho_{\infty}-\nabla\rho_{1}\cdot\nabla\Upsilon_{0}^{(2)}/\rho_{\infty}-\partial_{j}(v_{0i}^{\perp}v_{0j}^{\perp}\partial_{i}(1+\Upsilon_{0}^{(2)}))/q_{\infty}^{2}
\label{eq40}
\end{align}
where the source terms can be computed in terms of $\Upsilon_{0}^{(n)}$ and the incompressible flow solution. Eq. (\ref{eq40}) could be amenable to analytic solution for the circle, but it would be of limited utility, as solutions to full-potential equations are not preserved by conformal transformations.

\section{Kutta condition for the adjoint full-potential equations}
\label{sec:6}
Despite significant work, the issue of wake and Kutta conditions for compressible potential flows is by no means closed and a precise formulation is still lacking. In this section, we revisit the Kutta condition  from a continuous-adjoint perspective and outline a possible resolution of this problem. We first introduce a Lagrange-multiplier enforcement at the trailing edge, then determine the multiplier by stationarity with respect to circulation, and finally show that the resulting singular boundary forcing generates the Kutta functions through Green-function kernels.

The most detailed continuous formulation of the full-potential adjoint equations including wake and Kutta conditions can be found, to the best of our knowledge, in \cite{reuther1996} and \cite{kuruvila1995}. In both cases, the analysis is carried out in terms of the velocity potential and the issue of perturbations to the Kutta condition is approached by introducing a cut line, extending from the sharp trailing edge into the far field, across which the potential has a discontinuity proportional to the value of the circulation. The cut is considered as an interior boundary (similar to the treatment of adjoint equations with shocks \cite{giles2001}) across which certain consistency conditions need to be enforced. In ref. \cite{reuther1996}, the circulation is fixed, as a result of which the boundary integrals involving the cut lines cancel out, so the whole issue becomes irrelevant and is therefore not discussed further. In \cite{kuruvila1995}, on the other hand, the circulation is not fixed and integration along the cut results in the introduction of two delta functions with opposite signs centered at the trailing edge. This forcing structure does not reproduce the analytic adjoint solutions obtained
in \cite{lozano2025}  or the kernel structure derived below. 


More recently, Anevlavi and Belibassakis \cite{anevlavi2021} developed a continuous adjoint-based optimal shape design application for cavitating hydrofoils. They used a source-vorticity primal solver---which avoids a wake cut entirely---and imposed the adjoint Kutta condition as a local constraint term written in terms of the upper and lower tangential velocities at the trailing edge. The corresponding Lagrange multiplier was subsequently set to zero to cancel the resulting variational boundary terms, thereby removing the sensitivity to circulation from the adjoint field. We will follow a similar approach in section \ref{subsec:61} to enforce the Kutta condition, with several key differences that will allow us to recover the results obtained in section \ref{sec:3}.

\subsection{Imposing the Kutta condition} 

\label{subsec:61}
We will impose the Kutta condition by means of Lagrange multipliers, the value of which will be fixed by enforcing stationarity of the Lagrangians with respect to the circulation. The strategy of enforcing the adjoint Kutta condition by a trailing-edge penalty
term, with the multiplier fixed by stationarity with respect to circulation, was
proposed for the potential formulation in~\cite{letter}, where the resulting multiplier was reported without derivation. Here we give the full derivation for the full-potential adjoint in both the potential and stream-function formulations, and develop the
Green-function kernel representation of the resulting trailing-edge forcing
(Section~\ref{ssec:kernel}), which is compared against numerical adjoint solutions
in Section~\ref{sec:7}.

First, we need to identify the relevant form of the Kutta condition. Let $\delta v_{t}^{te}=\hat{t}\cdot\delta\vec{v}$ be the tangent component of the velocity perturbation at the trailing edge, where $\hat{t}$ is the unit tangent vector directed such that the profile is traveled in counter-clockwise sense. Near sharp trailing edges, $\hat{t}\cdot\delta\vec{v}$ behaves as \cite{coclici1999,bassanini1999}
\begin{equation}
(regular~terms)+(c_{0}+c_{1}\delta\Gamma)/r^{\beta}
\label{eq43}
\end{equation}
where $\delta\Gamma$ is the perturbation to the circulation (in conventions where a counter-clockwise vortex carries positive circulation), $r$ is the distance to the trailing edge, $\beta=(\pi-\tau)/(2\pi-\tau)$, where $0\le\tau\le\pi$ is the trailing-edge angle,  and $c_{0}$ and $c_{1}$ are constants. Notice that $0\le\beta\le 1/2$, the lower limit corresponding to blunt trailing edges and the upper limit corresponding to cusped trailing edges. The split coefficient of the singular term in eq. (\ref{eq43}) corresponds to the splitting of the perturbed solution near the trailing edge into a component with zero circulation and another with circulation $\delta\Gamma$ \cite{bassanini1999}. Both components are singular at the sharp trailing edge, with the same exponent, and the Kutta condition amounts to adjusting the circulation such that both singularities cancel out, i.e.
\begin{equation}
c_{0}+c_{1}\delta\Gamma=0
\label{eq44}
\end{equation}

Let us consider $r^{\beta}\delta v_{t}$ near the trailing edge. It is clear from (\ref{eq43}) that $r^{\beta}$ cancels the velocity singularity in such a way that $r^{\beta}\delta v_{t}^{te}\sim c_{0}+c_{1}\delta\Gamma$. Hence, we can impose the Kutta condition through the vanishing of the term
\begin{equation}
\lim_{r\rightarrow 0} r^{\beta} \delta {v}_{t} 
\label{eq42}
\end{equation}
In this form, (\ref{eq42}) is not very useful since it contains the distance to the trailing edge, a limit and its dependence on the circulation is given in terms of the constants $c_0$ and $c_1$ whose value is unknown. In the incompressible case \cite{lozano2025}, we used a conformal mapping transforming the airfoil with a sharp trailing edge into a circle. Near the trailing edge, the modulus of the mapping behaves as $h\sim Cr^{\beta}$ where $C$ is some constant. The quantity $h\hat{t}\cdot\delta\vec{v}$ in the airfoil plane maps to $\hat{t}\cdot\delta\vec{v}=(\hat{t}\cdot\delta\vec{v})_0+\delta\Gamma/2\pi$ in the circle plane, where the trailing edge is actually the rear stagnation point and $(\hat{t}\cdot\delta\vec{v})_0$ is the zero circulation part. So in that case, the dependence of the constraint term with respect to the circulation is explicit. 

In the compressible case, we will use a similar approach by converting the flow into an incompressible flow around a circle using Bers' map \cite{bers1954}. According to Bers, for any two-dimensional subcritical potential flow about an airfoil there exists a quasi-conformal mapping of the domain exterior to the airfoil to the exterior of a circle such that in the new coordinates the complex potential is holomorphic. Details of the quasi-conformal mapping, along with some useful formulae, are given in Appendix \ref{app:B}. Here, we limit ourselves to point out that, in terms of the quasi-conformal mapping $z\rightarrow\zeta$, where $z$ is the complex coordinate in the airfoil plane and $\zeta$ is the complex coordinate in the circle plane, we have
\begin{equation}
h_{te}\hat{t}\cdot\delta\vec{v}_{te}^{comp} = \frac{1+\tilde{\rho}}{2\tilde{\rho}}\partial_t\delta\phi^{inc}|_{rsp}
\label{eq47}
\end{equation}
where $h=|\partial z/\partial\zeta|$ is the modulus of the mapping (which, as in the conformal case, behaves as $r^{\beta}$ near sharp trailing edges), $\tilde{\rho}=\rho/\rho_\infty$ and $\delta\phi^{inc}$ is the perturbed (incompressible) potential in the circle plane. At the rear stagnation point of the circle we have \cite{lozano2025}
\begin{equation}
\partial_{t}\delta\phi^{inc}|_{rsp}=\delta\Gamma/2\pi+\partial_{t}\delta\phi_{\Gamma=0}^{inc}|_{rsp}
\label{eq48}
\end{equation}
where $\partial_{t}\delta\phi_{\Gamma=0}^{inc}|_{rsp}$ denotes the zero circulation part of the perturbed velocity. Notice that the dependence of the constraint term on the circulation is again explicit. 

Finally, we need to write the perturbed velocity in terms of the potential and stream function. For the former, we have 
\begin{equation}
\hat{t}\cdot\delta\vec{v}=\partial_{t}\delta\phi
\label{eqPot}
\end{equation}
while for the latter we recall that $\vec{v}=(\partial_{y}\psi,-\partial_{x}\psi)/\rho$. Using eq. (\ref{eq16}),

\begin{equation}
\delta\vec{v}
=
\rho^{-1}(\partial_y\delta\psi,-\partial_x\delta\psi)
-\vec{v}\,\delta\rho/\rho
=
\rho^{-1}(\partial_y\delta\psi,-\partial_x\delta\psi)
+a^{-2}\vec{v}(\vec{v}\cdot\delta\vec{v}) .
\label{eq52}
\end{equation}
%
%
%
%
Projecting both sides onto the tangent direction, using the flow tangency condition and rearranging yields
\begin{equation}
\hat t\cdot\delta\vec v
=
\rho^{-1}(1-M^2)^{-1}\partial_n\delta\psi .
\label{eq54}
\end{equation}

We now introduce Lagrange multipliers $\lambda^K_{\phi}$ and $\lambda^K_{\psi}$ to enforce the Kutta condition at the trailing edge. The augmented Lagrangians are
\begin{equation}
L =\int_{S}c_{\infty}^{-1}p(\hat{n}\cdot\vec{d})ds-\int_{\Omega}\tilde{\phi}\nabla\cdot(\rho\nabla\phi)d\Omega+\lambda^K_{\phi}h\partial_{t}\phi|_{te}
\label{eq71}
\end{equation}
and
\begin{equation}
L =\int_{S}c_{\infty}^{-1}p(\hat{n}\cdot\vec{d})ds-\int_{\Omega}\tilde{\psi}\nabla\cdot({\rho}^{-1}\nabla\psi)d\Omega+\lambda^K_{\psi}h\partial_{n}\psi|_{te}
\label{eq72}
\end{equation}
In (\ref{eq71}) and (\ref{eq72}), the quantities $h\partial_{t}\phi|_{te}$ and $h\partial_{n}\psi|_{te}$ have to be understood in the sense of (\ref{eq42}) i.e., as the limiting value at the trailing edge of the product of a vanishing quantity $h$ and a potentially singular quantity whose combined value is finite and given essentially by (\ref{eq44}). This interpretation will be kept in the remainder of the section and will be made explicit in the final results at the end of \ref{ssec:kernel}.  

Linearization of (\ref{eq71}) has to be done with care. When circulation is allowed to vary, the variational potential $\delta\phi$ is multivalued and requires special treatment. We decompose the variational potential as \cite{reuther1996}
\begin{equation}
\delta\phi=\delta\phi^{(0)}+\delta\Gamma\,\phi^{(\Gamma)},
\label{eqVarPot}
\end{equation}
where $\delta\phi^{(0)}$ is the single-valued zero-circulation part and $\phi^{(\Gamma)}$ is a
prescribed unit-circulation mode (the pull-back under Bers' mapping of the standard unit-circulation mode in the circle plane --- see Appendix \ref{app:B}).

Linearizing (\ref{eq71}) with respect to $\delta\phi$ using  \eqref{eqVarPot} yields 

\begin{align}
\delta L =&-\int_{S}c_{\infty}^{-1}\rho\nabla\phi\cdot\nabla\delta\phi^{(0)}(\hat{n}\cdot\vec{d})ds-\int_{\Omega}\tilde\phi\nabla\cdot\left(\hat{A}\nabla{\delta\phi^{(0)}}\right)d\Omega+\lambda^K_{\phi}h\partial_{t}\delta\phi^{(0)}|_{te} \nonumber\\ &-\delta\Gamma\int_{S}c_{\infty}^{-1}\rho\nabla\phi\cdot\nabla\phi^{(\Gamma)}(\hat{n}\cdot\vec{d})ds-\delta\Gamma\int_{\Omega}\tilde{\phi}\nabla\cdot\left(\hat{A}\nabla{\phi^{(\Gamma)}}\right)d\Omega+\delta\Gamma\lambda^K_{\phi}h\partial_{t}\phi^{(\Gamma)}|_{te}
\label{eqDecompPertLag}
\end{align}
Since Bers' mapping depends on the flow, the linearization of the Kutta constraint term also
generates the additional contribution
\begin{equation}
    \lambda_\phi^K\,\delta h\,\partial_t\phi|_{te}.
\end{equation}
This term vanishes for non-zero trailing-edge angle by the Kutta condition, and in the cusped case
because the original and perturbed moduli both vanish with exponents close to 1/2 at the trailing
edge~\cite{bers1954}.


The contribution of $\delta\phi^{(0)}$ in \eqref{eqDecompPertLag} can be integrated by parts on the closed wall contour and in the domain term as in section \ref{subsec:21} and yields the adjoint PDE and wall boundary condition (\ref{eq12}). The zero circulation part of the Kutta constraint term yields an additional contribution to the adjoint wall forcing proportional to the Kutta multiplier, as we shall see momentarily. The contribution of the circulation part, on the other hand, only depends on the flow perturbation through $\delta\Gamma$ and does not contribute to the adjoint equation or the wall forcing. The first term (the circulation contribution to the cost function) is zero for drag and 
\begin{equation}
-\delta\Gamma\int_{S}c_{\infty}^{-1}\rho\nabla\phi\cdot\nabla\phi^{(\Gamma)}(\hat{n}\cdot\vec{d})ds = -\rho_{\infty} q_\infty \delta\Gamma / c_{\infty}
\label{eqLiftfromCirculation}
\end{equation}
for lift \cite{finn1958}. The middle term is zero, since the circulation mode obeys the linearized potential equation, and the final term can be transformed to the circle plane and contributes to the computation of the Lagrange multiplier.  

The above treatment is only required for the variational potential. The variational stream function, on the other hand, is single-valued and the wall variation is handled through $\partial_n\delta\psi$, not through a tangential derivative of a multivalued potential. We can at any rate decompose the variational stream function into a zero circulation and a circulation part, but for $\psi$ that is just bookkeeping, not a topological necessity. Both parts are still single-valued, and the stream-function stationarity calculation can therefore be carried out directly.

Turning to our main goal---determining the Kutta multipliers $\lambda^K_{\phi}$ and $\lambda^K_{\psi}$ ---it turns out that, as explained in \cite{letter}, their value can be obtained by enforcing stationarity of the Lagrangians (\ref{eq71}) and (\ref{eq72}) with respect to circulation variations with all other quantities fixed (geometry, angle of attack, density, free-stream velocity), i.e., 
\begin{equation}
\frac{\delta L}{\delta\Gamma}=0
\label{eq73}
\end{equation}

From (\ref{eq73}) we obtain the following results. 
For drag, the corresponding multipliers vanish, whereas for lift, one obtains 
\begin{align}
\lambda^K_{\phi}
&=
\frac{4\pi \rho_\infty q_\infty}{c_\infty}
\frac{\tilde\rho_{te}}{1+\tilde\rho_{te}},
\nonumber \\
\lambda^K_{\psi}
&=
\frac{4\pi q_\infty}{c_\infty}
\frac{1}{(1-M_{te}^2)(1+\tilde\rho_{te})}.
\label{eq75}
\end{align}

The proof is as follows. Let $\cal{K}$ denote the Kutta constraint term. Then, noting that the circulation mode obeys the linearized potential equation, stationarity with respect to circulation amounts to 
\begin{align}
\frac{\partial I}{\partial\Gamma}+\frac{\partial \cal{K}}{\partial\Gamma}=0
\label{eqStationarityFull}
\end{align}
For drag, the objective function is independent of circulation, so $\partial I/\partial\Gamma=0$, and the corresponding multipliers vanish. For lift, one has, from (\ref{eqLiftfromCirculation}) (and the analogous relation for the stream function), $\partial I/\partial\Gamma = -\rho_{\infty} q_\infty / c_{\infty}$. 
Using this result, the trailing-edge relation \eqref{eq47}, the decomposition \eqref{eq48}, the potential
identity \eqref{eqPot} and the stream-function relation \eqref{eq54}, eq. (\ref{eqStationarityFull}) yields \eqref{eq75}. For the potential, the Kutta constraint term contains only the circulation part $\delta\Gamma\partial_{t}\phi^{(\Gamma)}|_{te}$. While Bers' map relations \eqref{eq47}--\eqref{eq48} and \eqref{eqPot} are written in terms of the full variational potential $\delta\phi$, the first-variation formulae \eqref{eq101}--\eqref{eq102} in Appendix \ref{app:B} are linear in the perturbation and thus apply in particular to
the unit-circulation mode $\phi^{(\Gamma)}$, whose image in the circle plane is the corresponding unit-circulation perturbation that carries the dependence on $\delta\Gamma$.


\subsection{Kernel interpretation}
\label{ssec:kernel}

The presence of the new terms in the Lagrangians  (\ref{eq71})  and (\ref{eq72}) alters the adjoint boundary conditions. For the stream function, the constraint term can be written as   
\begin{equation}
\int_S \lambda^K_{\psi}h\partial_{n}\psi  \delta(s-s_{te}) ds
\label{eq56}
\end{equation}
where $\delta(s)$ is the Dirac delta function and $s_{te}$ denotes the location of the trailing edge. In order to determine the contribution of  (\ref{eq56}) to the adjoint problem, we need to linearize it with respect to flow perturbations $\delta\psi$. Assuming that the geometry of the airfoil is held fixed, we obtain
\begin{equation}
\int_S (\lambda^K_{\psi}\delta h\partial_{n}\psi + \lambda^K_{\psi}h\partial_{n}\delta\psi ) \delta(s-s_{te}) ds
\label{eqLinStreamf}
\end{equation}
The term proportional to $\delta h$ vanishes as explained in the previous section, which leaves 
\begin{equation}
\frac{4\pi q_\infty}{c_\infty}
\int_S
\frac{1}{(1-M_{te}^2)(1+\tilde\rho_{te})}
h\,\partial_n\delta\psi\,\delta(s-s_{te})\,ds .
\label{eq56b}
\end{equation}

 where we have used (\ref{eq75}). Inserting eq. (\ref{eq56b}) into eq. (\ref{eq18}), we see that the effect of (\ref{eq56b}) is to add a Dirac delta function centered at the trailing edge as a boundary forcing term, i.e.,
\begin{equation}
\tilde{\psi}_S = (\text{regular terms}) + \frac{4\pi\rho_{\infty}q_{\infty}}{c_{\infty}} \frac{\tilde{\rho}}{1+\tilde{\rho}} h \delta(s-s_{te})
\label{eq57}
\end{equation}
where the product $h\delta(s-s_{te})$ is understood in the renormalized trailing-edge sense described
above, and made explicit in the limiting representations below.

For the potential, on the other hand, linearization of the constraint term yields 
\begin{equation}
\frac{4\pi\rho_{\infty}q_{\infty}}{c_{\infty}}\int_{S}\frac{\tilde{\rho}}{1+\tilde{\rho}}h\partial_{s}\delta\phi^{(0)}\delta(s-s_{te})ds
\label{eq65}
\end{equation}
Inserting eq. (\ref{eq65}) into eq. (\ref{eq10}) and integrating by parts along the profile, we find that the effect of eq. (\ref{eq65}) is to add the derivative of the Dirac delta function centered at the trailing edge as a boundary forcing term
\begin{equation}
\partial_{n}\tilde{\phi}_{S}=(regular~terms)+\frac{4\pi\rho_{\infty}q_{\infty}}{\rho c_{\infty}}\partial_{s}\left(\frac{\tilde{\rho}}{1+\tilde{\rho}}h\delta(s-s_{te})\right)
\label{eq66}
\end{equation}

where the derivative of the Dirac delta function must be understood in the distributional sense, and its discrete
counterpart arises through consistent integration over control volumes or elements. The singular boundary forcings (\ref{eq57}) and (\ref{eq66}) generate the Kutta functions. Our main result here is that the Kutta functions can be identified with the Poisson kernel of the linearized stream-function operator and the tangential derivative of the Neumann Green function of the linearized potential operator. The proof is as follows. 

For the stream-function formulation, the adjoint stream function obeys eq. (\ref{eq19}) with Dirichlet boundary conditions given by (\ref{eq57}) and a vanishing condition at the far field. The solution to this problem can be obtained as follows. Let $G_{D}$ be the solution to
\begin{equation}
\partial_{x_{i}}(A_{ij}\partial_{x_{j}}G_{D}(\vec{x},\vec{y}))=\delta(\vec{x}-\vec{y})
\label{eq60}
\end{equation}
where $\delta(\vec{x}-\vec{y})$ is the Dirac delta function and the boundary conditions are homogeneous, $G_{D}|_{\partial\Omega}=0$. $G_{D}$ is the Green's function corresponding to the Dirichlet problem associated with Eq. (\ref{eq19}). Multiplying both sides of equation (\ref{eq60}) by a function $\tilde{\psi}(\vec{x})$ obeying eq. (\ref{eq19}) and integrating over the domain yields, after integrating by parts twice on the left-hand side, the following boundary representation for $\tilde{\psi}(\vec{x})$ 
\begin{equation}
\tilde{\psi}(\vec{x})=\int_{S}\tilde{\psi}_{S}(\vec{y})P(\vec{x},\vec{y})ds_{\vec{y}}
\label{eq61}
\end{equation}
In eq. (\ref{eq61}),
\begin{equation}
P=n_{i}A_{ij}\partial_{j}G_{D}={\rho}^{-1}(1-M^{2})^{-1}\partial_{n}G_{D}
\label{eq62}
\end{equation}
(where $n$ is the normal vector to $S$) is the Poisson kernel for the generalized Laplacian (\ref{eq19}) \cite{gruter1982}. Inserting \eqref{eq57} into the boundary representation formula (\ref{eq61}) yields the adjoint stream function as 
\begin{align}
\tilde{\psi}(\vec{x})&=(\text{regular terms})+\int_{S}\frac{4\pi\rho_{\infty}q_{\infty}}{c_{\infty}} \frac{\tilde{\rho}}{1+\tilde{\rho}} h \delta(s-s_{te})P(\vec{x},\vec{y})ds_{\vec{y}}\nonumber\\ &=(\text{regular terms})+\frac{4\pi\rho_{\infty}q_{\infty}}{c_{\infty}} \frac{\tilde{\rho}_{te}}{1+\tilde{\rho}_{te}}\lim_{s\to s_{te}} h(\vec{y}(s))\,P(\vec{x},\vec{y}(s))
\label{eq63}
\end{align}
As a further check of eq. (\ref{eq63}), we note that as $\vec{x}$ approaches the boundary, the Poisson kernel approaches a Dirac delta function centered at $\vec{y}(s_{te})$ \cite{fabes1984}, in agreement with eq. (\ref{eq57}). 

Finally, comparing (\ref{eq63}) with the singular part of the lift-based adjoint stream function in (\ref{eq28}) allows the identification
\begin{equation}
\rho(\vec{x})\left(1+\Upsilon^{(2)}(\vec{x})\right) = \frac{4\pi\rho_{\infty}\tilde{\rho}_{te}}{1+\tilde{\rho}_{te}}\lim_{s\to s_{te}} h(\vec{y}(s))\,P(\vec{x},\vec{y}(s))
\label{eq64}
\end{equation}
in complete analogy with the incompressible case \cite{lozano2025}.

The adjoint potential obeys the linearized potential equation (\ref{eq11}) with wall boundary condition (\ref{eq66}). The solution of this problem can be represented in terms of the  Green's function $\hat G_N$ for the exterior
Neumann problem associated with $\hat{\cal L}=\nabla\cdot(\hat A\nabla)$, characterized by
$\hat{\cal L}_x\hat G_N(x,y)=\delta(x-y)$, homogeneous Neumann condition $n_i\hat{A}_{ij}\partial_j\hat{G}_N =0$ on $S$, and unit conormal flux at the far field. The Green's function is defined up to an additive
constant, which is irrelevant here since only its tangential derivative appears below. Using the standard boundary representation formula, the solution to the Neumann problem can be written as 
\begin{equation}
\tilde{\phi}(\vec{x})=-\int_{S}\partial_{i}\tilde{\phi}_{S}(\vec{y})n_{j}\hat{A}_{ij}\hat{G}_{N}(\vec{x},\vec{y})ds_{\vec{y}}=-\int_{S}\rho\partial_{n}\tilde{\phi}_{S}(\vec{y})\hat{G}_{N}(\vec{x},\vec{y})ds_{\vec{y}}
\label{eq68}
\end{equation}
up to an additive constant, where we have used $n_{i}\hat{A}_{ij}\partial_{j}\tilde{\phi}_{S}=\rho\partial_{n}\tilde{\phi}_{S}$. Inserting eq. (\ref{eq66}) into eq. (\ref{eq68}) and integrating by parts along the boundary yields 
\begin{equation}
\tilde{\phi}(\vec{x})=(regular~terms)+\frac{4\pi\rho_{\infty}q_{\infty}}{c_{\infty}}\frac{\tilde{\rho}_{te}}{1+\tilde{\rho}_{te}}\lim_{s\to s_{te}} h(\vec{y}(s))\,\partial_s\hat G_N(\vec{x},\vec{y}(s))
\label{eq69}
\end{equation}
The singular term in eq. (\ref{eq69}) can be compared with the singular term in the adjoint potential solution (\ref{eq24}), allowing the identification of the Kutta function in terms of the boundary tangent derivative of the Neumann Green's function 
\begin{equation}
\Upsilon^{(1)}(\vec{x})=4\pi\rho_{\infty}\frac{\tilde{\rho}_{te}}{1+\tilde{\rho}_{te}}\lim_{s\to s_{te}} h(\vec{y}(s))\,\partial_s\hat G_N(\vec{x},\vec{y}(s))
\label{eq70}
\end{equation}


The above equations give a streamlined approach to the problem of the Kutta condition in subcritical potential flows, and reduce seamlessly to the results reported in \cite{lozano2025} in the incompressible limit. The caveat is, of course, that the mapping depends on the flow solution and not just on the geometry and is not known beforehand. Besides, it does not seem possible to use eq. (\ref{eq64}) or (\ref{eq70}) to obtain closed-form expressions for the Kutta functions. However, the approach clearly identifies the relevant boundary conditions and offers a mathematical interpretation of the Kutta functions and clarifies several of their properties.

 First, they show that
the singular part of the solution is entirely localized at the trailing edge, while away
from that point the Kutta functions inherit the usual regularity properties of solutions of the corresponding linearized elliptic equations. Second, they make it possible to put on firmer grounds the frozen-coefficient analysis carried out in section \ref{sec:5}. 
The local behavior near the trailing edge should be given by the corresponding kernels of the frozen operator, 
which explains why at a rear stagnation point the leading singular structure agrees with the incompressible one, while
at a cusped trailing edge it is expected to differ only through the corresponding
Prandtl-Glauert-type anisotropic rescaling. Finally, the far-field behavior can likewise
be related to the Green's function of the operator frozen at the free stream state, leading to the
expected $O(1/r)$ decay and anisotropic stretching. In this sense, even though the Kutta functions are not known in
closed form, their representation in terms of Poisson kernels and derivatives of Green's
functions shows that they can be analyzed within the framework of elliptic kernel theory.


\subsection{Dual consistency and localized adjoint forcing}

The continuous adjoint Kutta formulation derived above is also relevant for dual
(adjoint) consistency, since a complete continuous adjoint problem---including its
boundary conditions---is a prerequisite for establishing that a discrete adjoint is
a consistent approximation of the continuous one, a property tied to optimal grid
convergence and to adjoint-based error estimation. The novelty in the present
setting is that the continuous adjoint problem contains a singular contribution
localized at the trailing edge, eqs.~\eqref{eq57} and~\eqref{eq66},
which is obtained \emph{without} incorporating the wake cut into the variational
formulation, in contrast to traditional treatments that enforce the Kutta condition
through non-local wake-cut constraints. A dual-consistent discretization must
therefore reproduce this singular trailing-edge forcing in the distributional sense,
in addition to the regular adjoint boundary-value problem. The implications of this
viewpoint for discrete adjoint consistency, including formulations without wake cuts
such as~\cite{anevlavi2021}, are discussed in more detail
in~\cite{letter}.

\section{Sample solutions}
\label{sec:7}
\subsection{Subcritical flow past an airfoil at $M_{\infty}=0.2$}
We will consider steady inviscid compressible shock-free irrotational flow past an airfoil. We will use a numerical adjoint solution to investigate the properties of the corresponding full-potential solution. We pick a symmetric van de Vooren airfoil given by the conformal transformation \cite{lozano2022,lozano2023}
\begin{equation}
z(\zeta)=1+\frac{(\zeta-R)^{k}}{(\zeta-\sigma R)^{k-1}}
\label{eq79}
\end{equation}
at zero incidence and free-stream Mach number $M_{\infty}=0.2$. Eq. (\ref{eq79}) defines the airfoil in the complex $z$ plane in terms of a circle of radius $R=(1+\sigma)^{k-1}/2^{k}$ centered at the origin in the complex $\zeta$ plane. Here, $\sigma=0.0371$ and $k=86/45$, resulting in a symmetric airfoil with 12\% thickness and finite trailing-edge angle $\tau=16^{\circ}$. The numerical solution is obtained with the SU2 solver \cite{economon2016} on an unstructured mesh with 6400 nodes on the airfoil profile and $1.54\times10^{6}$ nodes and $3.06\times10^{6}$ triangular elements throughout the flowfield, with an outer freestream boundary domain of around 100 chord lengths from the geometry. 

\begin{figure}[H]
\centering
    \begin{subfigure}[b]{0.5\textwidth}            
            \includegraphics[width=\textwidth]{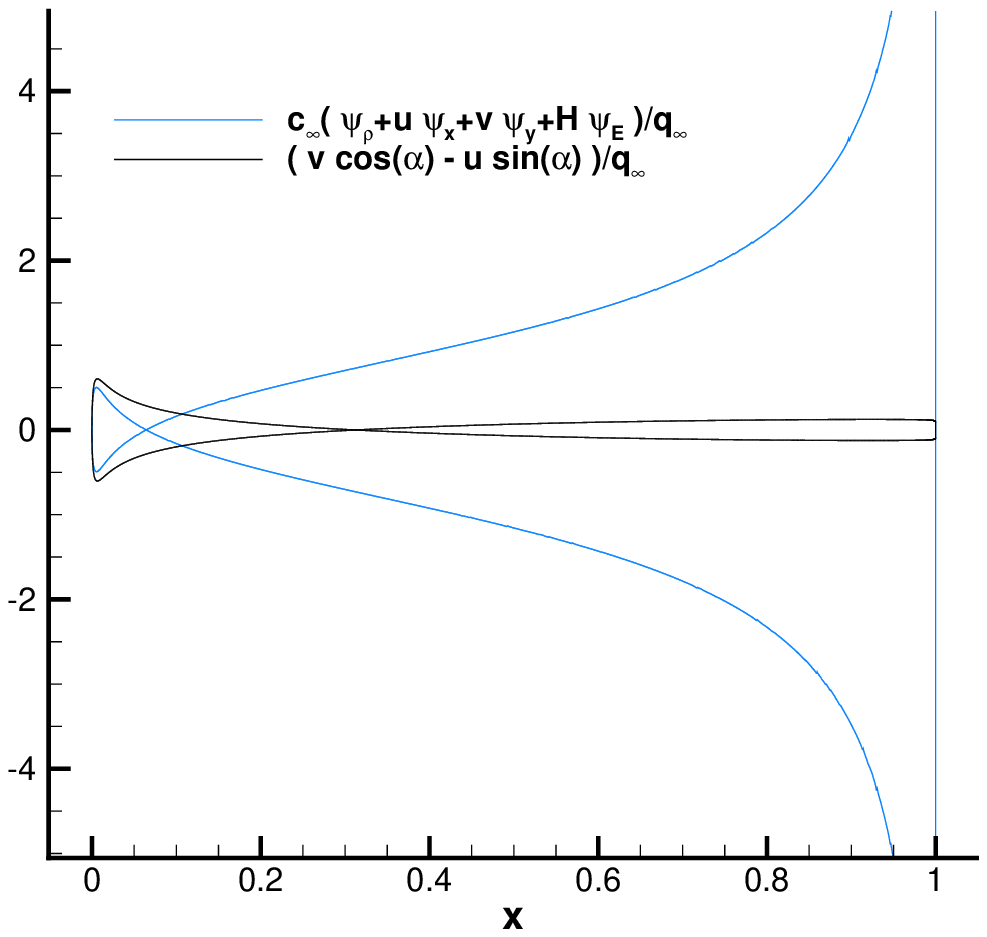}
    \end{subfigure}%
    \begin{subfigure}[b]{0.5\textwidth}
            \centering
            \includegraphics[width=\textwidth]{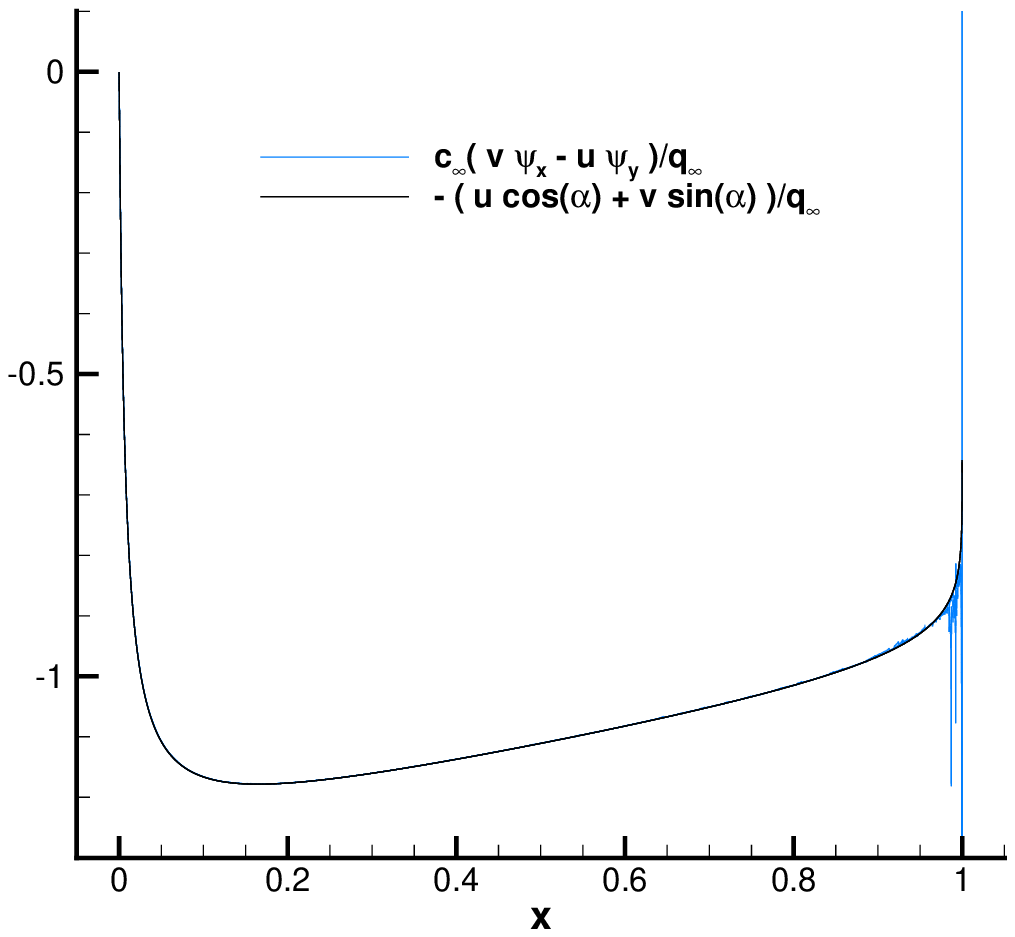}
    \end{subfigure}
    \caption{Lift-based full-potential adjoint solutions on a van de Vooren airfoil profile for subcritical flow with $M_{\infty}=0.2$ and $\alpha=0^{\circ}$ estimated from an adjoint Euler solution computed with the SU2 solver.}
    \label{fig:fig1}
\end{figure}

Using Eq. (\ref{eq32}), we can estimate the adjoint full-potential solutions from the numerical flow and adjoint compressible Euler solution. This is shown in Figure \ref{fig:fig1} for the lift-based adjoint, which also shows the regular part of the adjoint solutions (\ref{eq35}). These are reconstructed scalar full-potential quantities from the Euler adjoint, not independently solved full-potential adjoints.

For the adjoint stream function, the numerical and regular analytic curves are nearly
indistinguishable except in a small neighborhood of the trailing edge, where the numerical
solution develops the expected singular structure. Since the difference between the two
curves is precisely $1+\Upsilon^{(2)}$, this provides numerical support for a
trailing-edge singularity in $\Upsilon^{(2)}$. In the incompressible limit, the analogous
contribution reduces to a Dirac-delta singularity.

\begin{figure}[H]
    \centering
    \includegraphics[width=0.6\textwidth]{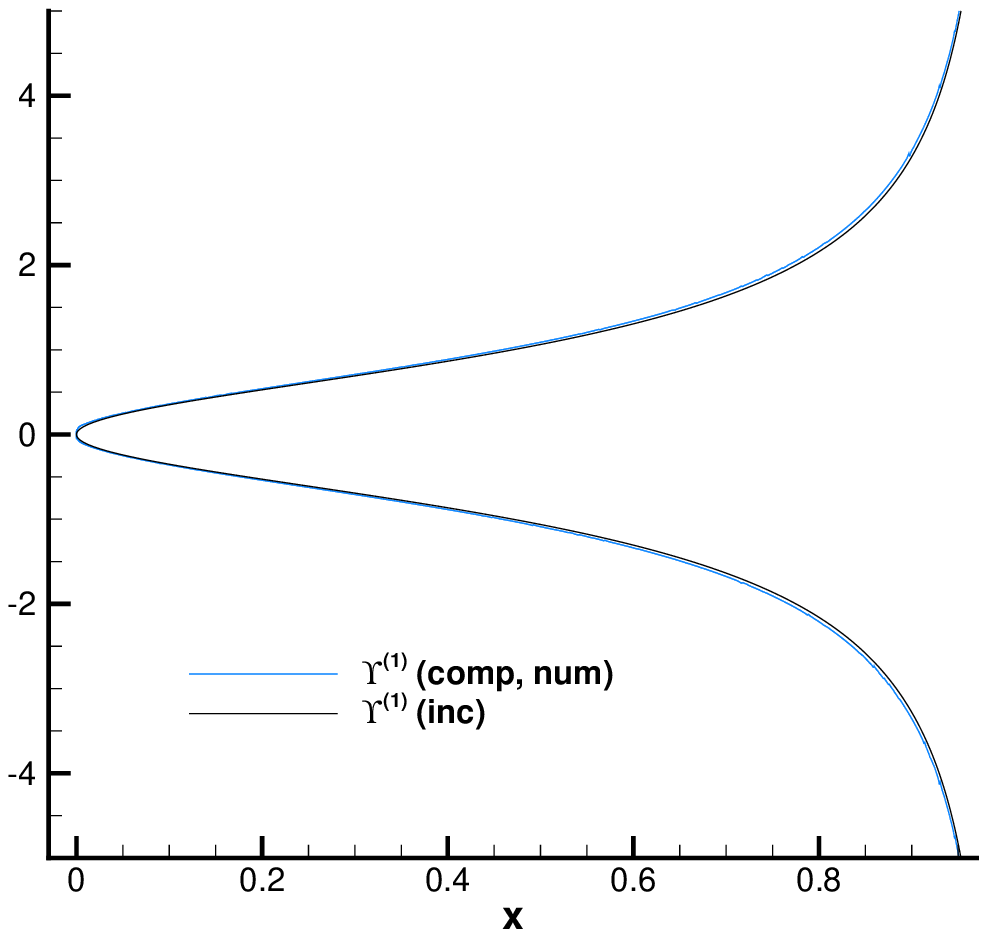}
    \caption{Difference between the numerical estimation of the adjoint potential and the regular part of the analytic solution compared with the incompressible $\Upsilon^{(1)}$ at the airfoil profile $M_{\infty}=0.2$ and $\alpha=0^{\circ}$.}
    \label{fig:fig2}
\end{figure}

For the adjoint potential, the difference between the numerical solution and the regular
analytic contribution isolates $\Upsilon^{(1)}$. This quantity is shown in Figure \ref{fig:fig2} together
with its incompressible counterpart. At $M_\infty=0.2$, the two are very close over most
of the profile, indicating that compressibility effects remain weak in this regime.

%
%

\begin{figure}[H]
    \centering
    \begin{subfigure}[b]{0.5\textwidth}            
            \includegraphics[width=\textwidth]{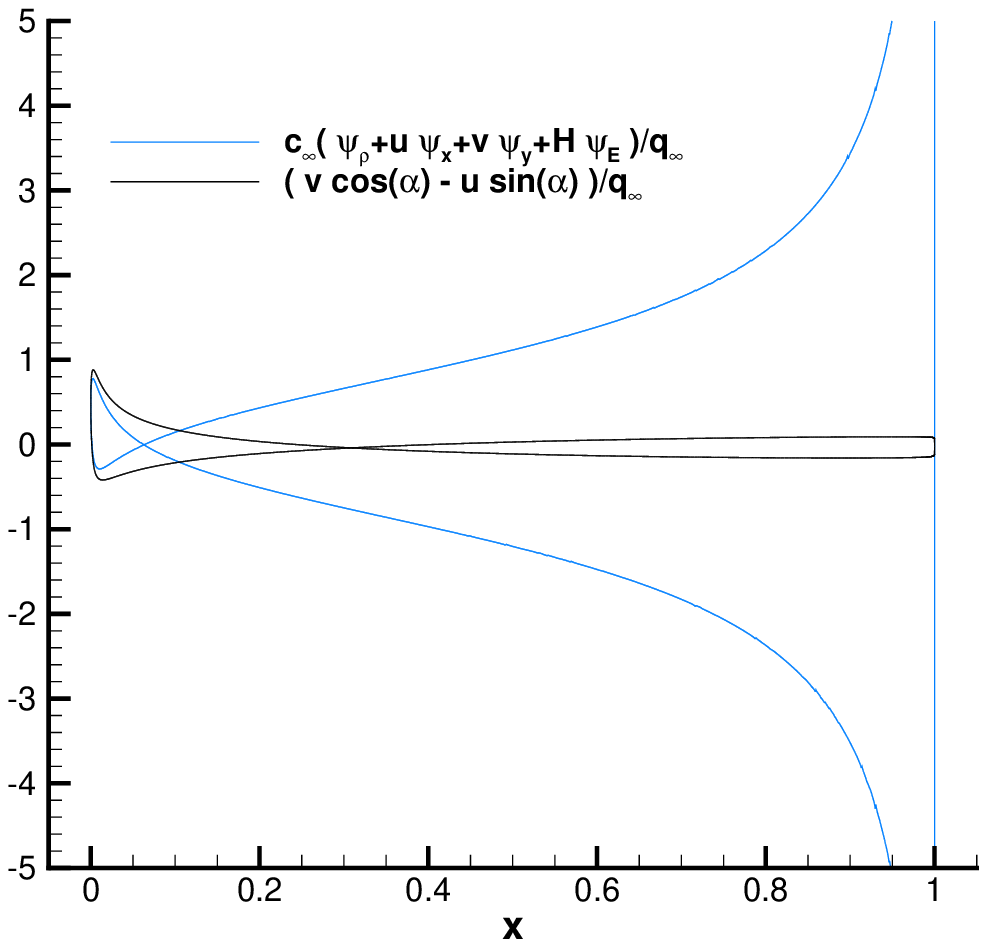}
    \end{subfigure}%
    \begin{subfigure}[b]{0.5\textwidth}
            \centering
            \includegraphics[width=\textwidth]{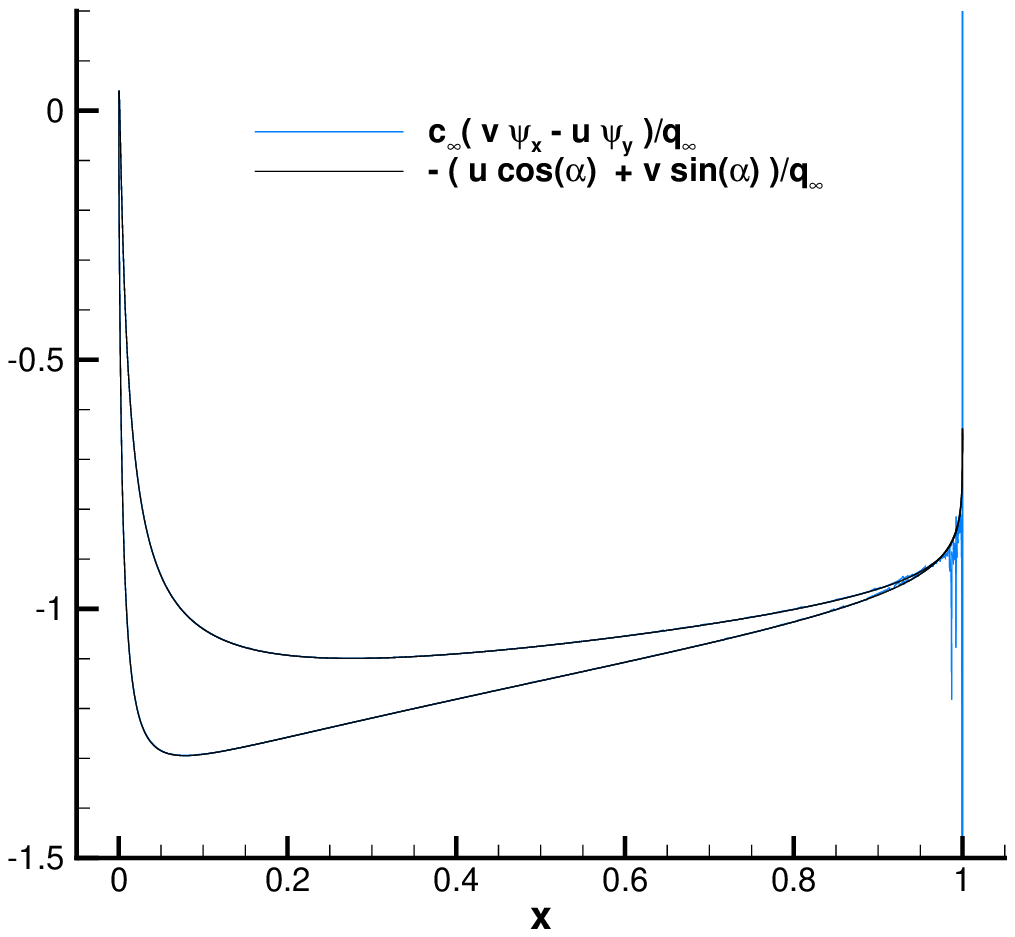}
    \end{subfigure}
    \caption{Lift-based full-potential adjoint solutions on a van de Vooren airfoil profile for subcritical flow with $M_{\infty}=0.2$ and $\alpha=2^{\circ}$ estimated from an adjoint Euler solution computed with the SU2 solver.}
    \label{fig:fig3}
\end{figure}

\begin{figure}[H]
    \centering
    \includegraphics[width=0.6\textwidth]{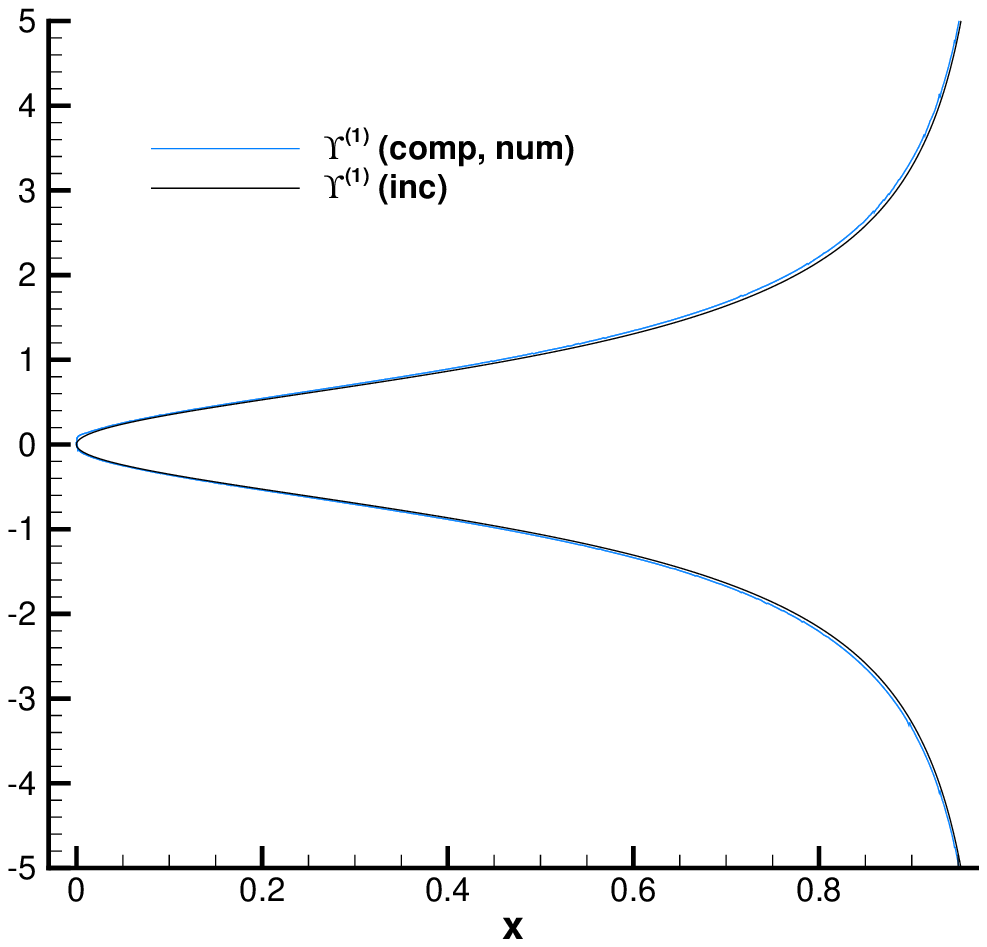}
    \caption{Difference between the numerical estimation of the lift-based adjoint potential and the regular part of the analytic solution compared with the incompressible $\Upsilon^{(1)}$ at the airfoil profile for a lifting solution with $M_{\infty}=0.2$ and $\alpha=2^{\circ}$}
    \label{fig:fig4}
\end{figure}

The same comparison can be carried out for a lifting base flow. The corresponding results,
shown in Figures \ref{fig:fig3} and \ref{fig:fig4}, confirm that the same decomposition remains valid away from the
trailing-edge singularity.

\subsection{Subcritical flow past an airfoil at $M_{\infty}=0.5$}

\begin{figure}[H]
    \centering
    \begin{subfigure}[b]{0.5\textwidth}            
            \includegraphics[width=\textwidth]{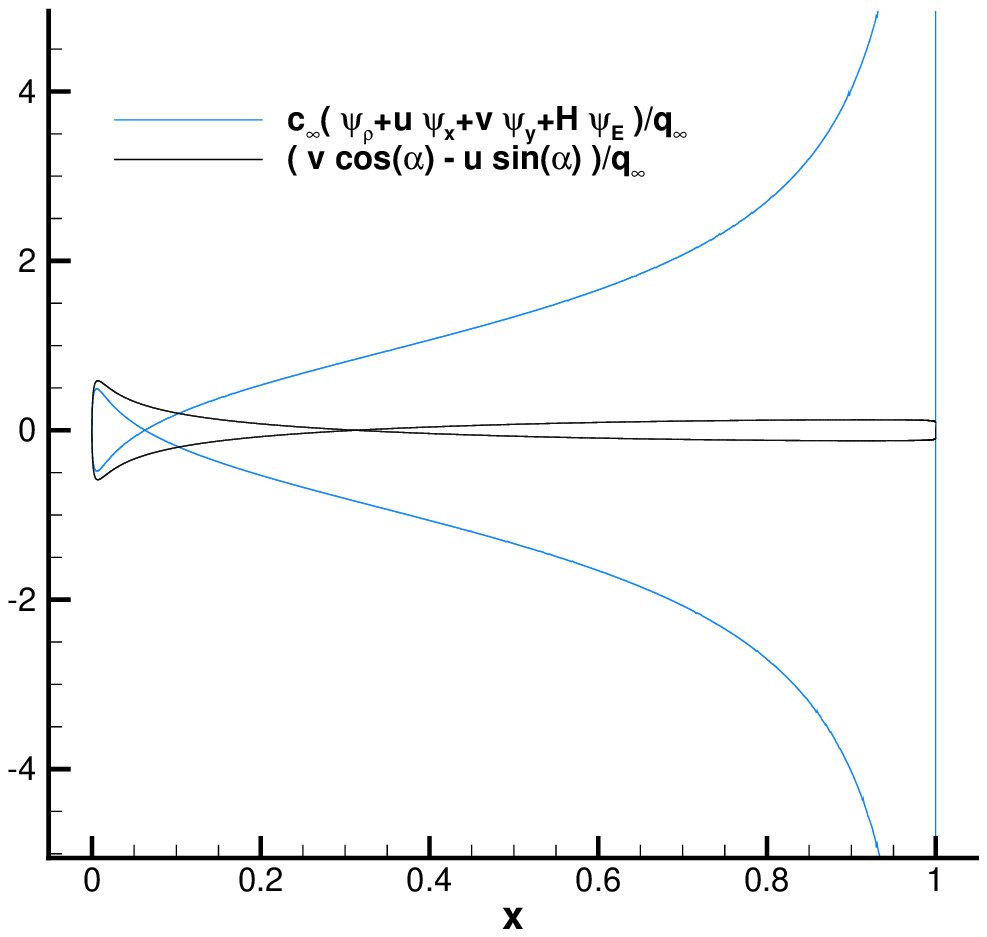}
    \end{subfigure}%
    \begin{subfigure}[b]{0.5\textwidth}
            \centering
            \includegraphics[width=\textwidth]{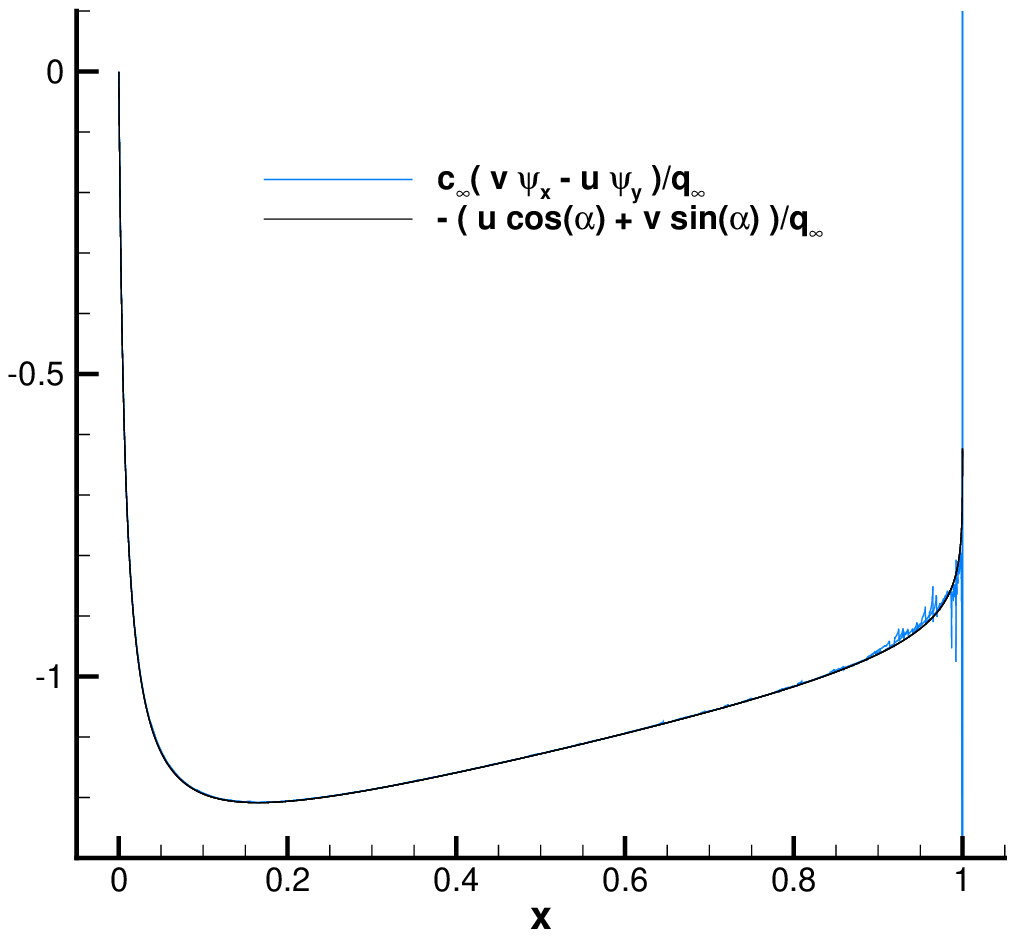}
    \end{subfigure}
    \caption{Lift-based full-potential adjoint solutions on a van de Vooren airfoil profile for subcritical flow with $M_{\infty}=0.5$ and $\alpha=0^{\circ}$ estimated from an adjoint Euler solution computed with the SU2 solver.}
    \label{fig:fig7}
\end{figure}

\begin{figure}[H]
    \centering
    \includegraphics[width=0.6\textwidth]{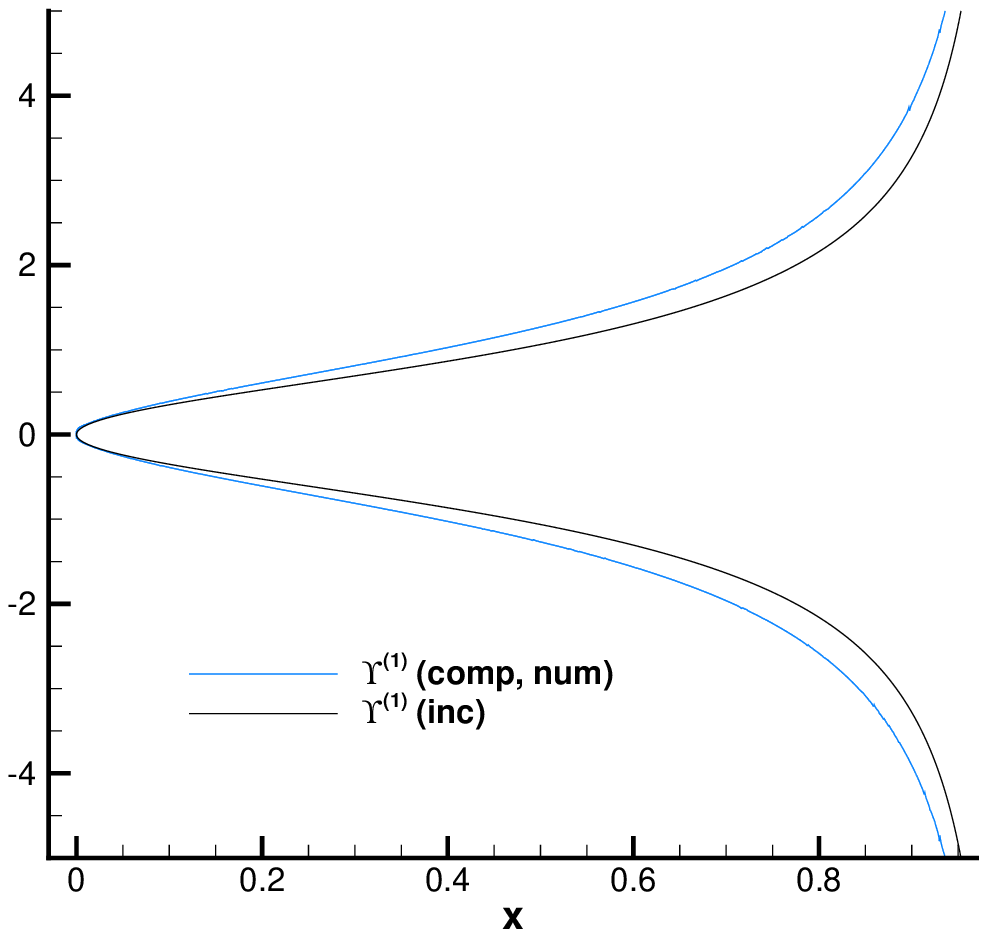}
    \caption{Difference between the numerical estimation of the adjoint potential and the regular part of the analytic solution compared with the incompressible $\Upsilon^{(1)}$ at the airfoil profile for $M_{\infty}=0.5$ and $\alpha=0^{\circ}$.}
    \label{fig:fig8}
\end{figure}

At the higher free-stream Mach number $M_\infty=0.5$ (which is still uniformly subsonic throughout the domain), the same qualitative structure is
observed in Figures \ref{fig:fig7} and \ref{fig:fig8}. The adjoint stream function continues to agree closely with the regular analytic
prediction except in the immediate vicinity of the trailing edge, where the singular
contribution dominates. The singular part of the adjoint potential, shown in
Figure \ref{fig:fig8}, still has the same overall shape as in the lower-Mach case, but now departs
noticeably from the incompressible $\Upsilon^{(1)}$. This provides a clearer indication of
compressibility effects in the Kutta correction.

The comparisons in Figures~\ref{fig:fig1}--\ref{fig:fig8} validate the Euler--full-potential
correspondence \eqref{eq32} and the decomposition of the adjoint fields into a regular part and
a trailing-edge singular part, from which $\Upsilon^{(1)}$ and $\Upsilon^{(2)}$ are extracted
along the wall. The close agreement of the extracted $\Upsilon^{(1)}$ with the closed-form
incompressible kernel at $M_\infty=0.2$, together with the structured growth of the deviation at
$M_\infty=0.5$, is consistent with the kernel representation of Section~\ref{ssec:kernel}, in
which the compressible kernel reduces to its incompressible counterpart as $M\to0$ and
compressibility enters through the operator coefficients. It should be stressed, however, that
the Kutta functions are recovered here through the Euler adjoint rather than by solving the forced
linearized full-potential problem; an independent solution of that forced problem, which would
directly exercise the singular trailing-edge forcings \eqref{eq57} and \eqref{eq66} and the kernel
identifications \eqref{eq64} and \eqref{eq70}, lies beyond the scope of the present paper.

\section{Conclusions}

In this paper, we have examined a number of structural properties of the adjoint
full-potential equations for two-dimensional subcritical flow. After reviewing the adjoint
equations in both the potential and stream-function formulations, we have used the
Green's-function viewpoint and the connection with the compressible adjoint Euler equations
to identify the adjoint potential and stream function with the responses to point mass and
vorticity sources. This provides a natural interpretation of the two adjoint variables and
clarifies their relation to the corresponding Euler adjoint variables.

For lift-based cost functions, the resulting adjoint fields contain two unknown Kutta
functions that encode the circulation correction required to preserve the Kutta condition.
Although these functions are not available in closed form in the compressible case, we have
shown that they obey the linearized full-potential operators and are linked by a pair of
differential relations that generalize the Cauchy-Riemann equations. In the incompressible
limit, they reduce to the Poisson kernel for the Laplacian on the exterior of the circle and
its harmonic conjugate. In this sense, the compressible theory preserves the operator and
kernel structure of the incompressible case even though the explicit formulas are lost.

A second objective of the paper has been to clarify the role of the Kutta condition in the
continuous adjoint full-potential problem. We have shown that the Kutta condition may be
incorporated by adding to the Lagrangian a constraint term imposing the vanishing of the singular part of 
the velocity at the trailing edge without explicit use of a cutting line. The value of the Lagrange multiplier can be obtained by demanding stationarity of the Lagrangian with respect to circulation. The effect of the constraint term is to introduce renormalized point-supported boundary forcings in the adjoint wall boundary conditions, represented formally by a Dirac delta
function and its tangential derivative. The corresponding Green-function representations identify
the Kutta functions with boundary derivatives of the Green functions of the linearized
operators.

The numerical results support this picture. At low Mach number, the extracted Kutta
function remains close to its incompressible counterpart, whereas at higher subcritical Mach
number the same decomposition persists but the deviation from the incompressible kernel
becomes appreciable. This is consistent with the view advocated here: the compressible case
does not retain the explicit formulas of the incompressible theory, but it does retain a
clear structural framework in terms of linearized operators, generalized conjugacy, Green
functions, and continuous Kutta enforcement.

In addition to their analytical interest, the results presented here identify the singular
boundary structure---the kernel representation of the Kutta functions and the renormalized
trailing-edge forcing---that a consistent discretization of the continuous adjoint problem should
reproduce. The numerical comparisons reported here provide indirect support for this picture; a
direct solution of the forced full-potential adjoint problem, as opposed to the reconstruction via
the Euler adjoint used here, remains an open task and would itself furnish a natural verification
benchmark for full-potential adjoint solvers, as well as a test bed for adjoint-based error
estimation and design methods for full-potential flow.

Although the paper focuses on force functionals for definiteness, the same construction extends to other outputs, such as moments, provided that the derivative of the output with respect to the circulation mode can be identified. The explicit representation used here relies on quasi-conformal mappings. This is specific to the present two-dimensional analysis and serves mainly to make the Kutta terms and their dependence on circulation explicit; a fully physical-plane formulation would be preferable for numerical purposes, but is not pursued here.

The present paper also establishes a connection with the adjoint Euler equations, but its scope should not be overstated. The results presented here resolve only the elliptic/potential part of the singular adjoint Euler structure,
namely the sector carried by the finite wall combinations $I^{(1)}$ and $I^{(2)}$. The link with the Euler equations is thus exact in this restricted sense, but it cannot be directly extrapolated to the full Euler setting.  In particular, the Kutta
multiplier construction of Section~\ref{sec:6} uses the scalar circulation degree of freedom
available in potential flow and should not be interpreted as an analogous point-constraint formulation for the general Euler adjoint equations. The full Euler lift-adjoint solution \eqref{eq33} couples the singular
contributions associated with $I^{(1)}$ and $I^{(2)}$, and also includes
additional non-potential components, in particular the streamline-integral term \(\Xi\).
These non-potential terms are responsible for singular behavior that is not captured by the
full-potential formulation, including the singular structure associated with the incoming
stagnation streamline and the wall. Understanding the continuous adjoint formulation
corresponding to those structures is a separate problem and is not addressed here.


\begin{appendices}
\section{Lift and drag of the compressible point source and vortex}
\label{app:A}




Let us compute the force exerted by the Green's functions. By this we mean, of course, the additional force exerted on the profile when the base flow is perturbed by a point source. This result was obtained in \cite{lozano2023} and we will briefly summarize it here for completeness. The computation is based on the analytic adjoint solution for the drag-based adjoint equations  (\ref{eq80}) obtained in \cite{lozano2021}. 
This solution is valid for two-dimensional inviscid, isentropic and irrotational flows (such as, for example, subcritical inviscid flow past 2D airfoils). It was claimed in \cite{lozano2021} that this solution is only valid for non-lifting flows because for lifting flows eq. (\ref{eq80}) does not tend to zero sufficiently fast. This conclusion is wrong, as was subsequently demonstrated in \cite{lozano2022} for the incompressible case, where it was argued that the far-field boundary condition does not require the adjoint state to vanish as $(1/r^{2})$ in the far field but only that $\int_{S_{\infty}}\Psi^{T}\hat{n}\cdot\vec{ A}\delta Uds=0$, which is verified in the lifting case as well. A similar argument can be used for the compressible solution (\ref{eq80}).

As explained in \cite{giles1997}, it is possible to compute the linearized drag functionals for the mass source and normal point force using eq. (\ref{eq80}) as 
\begin{equation}
\begin{pmatrix} \delta I_{D}^{(1)} \\ \delta I_{D}^{(2)} \end{pmatrix} = \begin{pmatrix} 1 & u & v & H \\ 0 & -\rho v & \rho u & 0 \end{pmatrix} \Psi_{D} = \frac{1}{c_{\infty}q_{\infty}} \begin{pmatrix}(\vec{q}-\vec{q}_{\infty})\cdot\vec{q}_{\infty} \\ \rho (\vec{q}-\vec{q}_{\infty})^{\perp}\cdot\vec{q}_{\infty} \end{pmatrix}
\label{eq81}
\end{equation}
where $\vec{q}^{\perp}\equiv(-v,u)$. 
By analogy with the incompressible case, the forces exerted on the wall
\emph{per unit source strength} by the mass source and the vortex are,
respectively,
\begin{equation}
\vec F_1 = \vec q - \vec q_\infty, \qquad
\vec F_2 = \rho\,(\vec q - \vec q_\infty)^{\perp},
\label{eq82}
\end{equation}
the mass source being normalized by its mass flux and the vortex by its circulation. The two reference strengths differ by a
density, which is why $\vec F_2$ carries an explicit factor $\rho$ while
$\vec F_1$ does not. 

Using eq. (\ref{eq82}), it is possible to write the linearized lift functionals as
\begin{align}
\delta I_{L}^{(1)}&=\frac{1}{c_{\infty}}\left(\vec{F}_{1}\cdot\hat{q}_{\infty}^{\perp}+q_{\infty}\Upsilon^{(1)}\right)=\frac{1}{c_{\infty}q_{\infty}}(vu_{\infty}-uv_{\infty})+\frac{q_{\infty}}{c_{\infty}}\Upsilon^{(1)} \nonumber 
\\
\delta I_{L}^{(2)}&=\frac{1}{c_{\infty}}\left(\vec{F}_{2}\cdot\hat{q}_{\infty}^{\perp}-\rho q_{\infty}\Upsilon^{(2)}\right)=\frac{\rho}{c_{\infty}q_{\infty}}(\vec{q}-\vec{q}_{\infty})\cdot\vec{q}_{\infty}-\rho\frac{q_{\infty}}{c_{\infty}}\Upsilon^{(2)}
\label{eq83}
\end{align}
where $\Upsilon^{(1)}$ and $\Upsilon^{(2)}$ are two unknown functions that incorporate the singular behavior of the adjoint solution at the trailing-edge/rear stagnation point. Physically, they represent the amount of circulation required to restore the Kutta condition in the presence of the corresponding point source, and they reduce, in the incompressible case, to the Poisson kernel for the circle and its harmonic conjugate.

\section{Quasi-conformal mapping for compressible potential flows}
\label{app:B}
For two-dimensional compressible potential flows, it is possible to define a quasi-conformal mapping from the airfoil plane to the exterior of a unit circle such that the complex potential is an analytic function on the circle plane and, thus, corresponds to the complex potential of an incompressible flow \cite{bers1954}. If $z=x+iy$ is the complex coordinate in the airfoil plane and $\zeta=\xi+i\eta$ is the complex coordinate in the circle plane, this means that the complex potential $\Phi=\phi+i\psi/\rho_\infty$, which is not an analytic function of $z$, is an analytic function of $\zeta$ and, in fact,
\begin{equation}
\Phi^{comp}(z)=\Phi^{inc}(\zeta(z))=A\left(e^{-i\alpha}\zeta(z)+\frac{e^{i\alpha}}{\zeta(z)}\right)+\frac{\Gamma}{2\pi i}\log \zeta(z)
\label{eq84}
\end{equation}
where $\alpha$, A and $\Gamma$ are real constants. Unlike the incompressible case, the mapping $\zeta(z)$ depends on the flowfield and is also present when the original profile is a circle, in which case it simply accounts for the compressibility in the flowfield. For a given flowfield, the mapping is uniquely determined by demanding that $\zeta(\infty)=\infty$ and $\zeta(z_{te})=1$ (resulting in what Bers calls the normalized mapping). 

In terms of this mapping, the velocity can be computed as
\begin{align}
\phi_{x}&=\xi_{x}\phi_{\xi}+\eta_{x}\phi_{\eta} \nonumber \\
\phi_{y}&=\xi_{y}\phi_{\xi}+\eta_{y}\phi_{\eta}
\label{eq85}
\end{align}
and an analogous relation for the stream function. 

As functions of \(\zeta=\xi+i\eta\), the potential and the normalized stream function
\(\psi/\rho_\infty\) obey the Cauchy-Riemann equations,
\[
\partial_\xi\phi=\rho_\infty^{-1}\partial_\eta\psi,
\qquad
\partial_\eta\phi=-\rho_\infty^{-1}\partial_\xi\psi .
\]
Demanding that in the airfoil plane
\[
\partial_x\phi=\rho^{-1}\partial_y\psi,
\qquad
\partial_y\phi=-\rho^{-1}\partial_x\psi
\]
imposes the following relationship between the derivatives of the mapping functions:
\begin{equation}
 \begin{pmatrix} \xi_{x} \\ \xi_{y} \end{pmatrix} = \begin{pmatrix} -\frac{\left(\rho^2-{\rho_{\infty}}^2 \right){\phi_{\xi}}\phi_{\eta}}{{\rho_{\infty}}^{2}{\phi_{\eta}}^{2}+\rho^{2}{\phi_{\xi}}^{2}} & \frac{\rho\rho_{\infty}\left({\phi_{\xi}}^{2}+{\phi_{\eta}}^{2}\right)}{{\rho_{\infty}}^{2}{\phi_{\eta}}^{2}+\rho^{2}{\phi_{\xi}}^{2}} \\ -\frac{\rho\rho_{\infty}\left({\phi_{\xi}}^{2}+{\phi_{\eta}}^{2}\right)}{{\rho_{\infty}}^{2}{\phi_{\eta}}^{2}+\rho^{2}{\phi_{\xi}}^{2}} & -\frac{\left(\rho^2-{\rho_{\infty}}^2 \right){\phi_{\xi}}\phi_{\eta}}{{\rho_{\infty}}^{2}{\phi_{\eta}}^{2}+\rho^{2}{\phi_{\xi}}^{2}}  \end{pmatrix}  \begin{pmatrix} \eta_{x} \\ \eta_{y} \end{pmatrix} 
 \label{eq86}
\end{equation}
such that (\ref{eq85}) can be written as
\begin{align}
\phi_{x}&=\xi_{x}\phi_{\xi}+\eta_{x}\phi_{\eta}=\rho_{\infty}\frac{{\phi_{\xi}}^{2}+{\phi_{\eta}}^{2}}{{\rho_{\infty}}^{2}{\phi_{\eta}}^{2}+\rho^{2}{\phi_{\xi}}^{2}}(\rho_{\infty}\phi_{\eta}\eta_{x}+\rho\phi_{\xi}\eta_{y}) \nonumber \\
\phi_{y}&=\xi_{y}\phi_{\xi}+\eta_{y}\phi_{\eta}=\rho_{\infty}\frac{{\phi_{\xi}}^{2}+{\phi_{\eta}}^{2}}{{\rho_{\infty}}^{2}{\phi_{\eta}}^{2}+\rho^{2}{\phi_{\xi}}^{2}}(-\rho\phi_{\xi}\eta_{x}+\rho_{\infty}\phi_{\eta}\eta_{y})
\label{eq87}
\end{align}
It is easy to check that, with the above, the complex velocity can be written as
\begin{equation}
w(z)=\phi_{x}-i\phi_{y}=2(1+\tilde{\rho})^{-1}\partial_{z}\zeta\partial_{\zeta}\Phi
\label{eq88}
\end{equation}
where
\begin{equation}
\partial_{z}\zeta=(\zeta_{x}-i\zeta_{y})/2=(\xi_{x}+i\eta_{x}-i\xi_{y}+\eta_{y})/2
\label{eq89}
\end{equation}
and
\begin{equation}
\partial_{\zeta}\Phi=(\Phi_{\xi}-i\Phi_{\eta})/2=\phi_{\xi}-i\phi_{\eta}=(\psi_{\eta}+i\psi_{\xi})/\rho_\infty
\label{eq90}
\end{equation}

The Jacobian determinant of the mapping is
\begin{equation}
J=\begin{vmatrix}\xi_{x}&\xi_{y}\\ \eta_{x}&\eta_{y}\end{vmatrix}=\xi_{x}\eta_{y}-\xi_{y}\eta_{x}=\frac{({\eta_{x}}^{2}+{\eta_{y}}^{2})\rho\rho_{\infty}({\phi_{\eta}}^{2}+{\phi_{\xi}}^{2})}{{\rho_{\infty}}^{2}{\phi_{\eta}}^{2}+\rho^{2}{\phi_{\xi}}^{2}}
\label{eq91}
\end{equation}
Likewise,
\begin{equation}
\xi_{x}^{2}+{\xi_{y}}^{2}=\frac{\rho^{2}{\phi_{\eta}}^{2}+{\rho_{\infty}}^{2}{\phi_{\xi}}^{2}}{{\rho_{\infty}}^{2}{\phi_{\eta}}^{2}+\rho^{2}{\phi_{\xi}}^{2}}({\eta_{x}}^{2}+{\eta_{y}}^{2})
\label{eq92}
\end{equation}
and
\begin{equation}
|\partial_{z}\zeta|^{2}=\frac{({\eta_{x}}^{2}+{\eta_{y}}^{2})(\rho+\rho_{\infty})^{2}({\phi_{\eta}}^{2}+{\phi_{\xi}}^{2})}{4({\rho_{\infty}}^{2}{\phi_{\eta}}^{2}+\rho^{2}{\phi_{\xi}}^{2})}=\frac{(\rho+\rho_{\infty})^{2}}{4\rho\rho_{\infty}}J
\label{eq93}
\end{equation}
which yields the following result for the modulus of the mapping
\begin{equation}
h=|\partial_{\zeta}z|=\frac{|\partial_{z}\zeta|}{J}=\frac{(\rho+\rho_{\infty})}{2\sqrt{\rho\rho_{\infty}}}J^{-1/2}
\label{eq94}
\end{equation}
When $\rho=\rho_{\infty}$, eq. (\ref{eq94}) reduces to $h=J^{-1/2}$ which is the correct relation in the conformal case where $\xi_{x}=\eta_{y}$ and $\xi_{y}=-\eta_{x}$.

Recall that the mapping $\zeta$ converts an airfoil into a unit circle. If the circle is parameterized by the polar angle $\theta$ in the $\zeta$ plane (such that the rear stagnation point of the circle $\zeta=1$ corresponds to $\theta=0$), the tangent vector in the airfoil plane is
\begin{align}
x_{\theta}&=-x_{\xi}\sin \theta+x_{\eta}\cos \theta=-\frac{1}{J}\eta_{y}\sin \theta-\frac{1}{J}\xi_{y}\cos \theta \nonumber \\
y_{\theta}&=-y_{\xi}\sin \theta+y_{\eta}\cos \theta=\frac{1}{J}\eta_{x}\sin \theta+\frac{1}{J}\xi_{x}\cos \theta
\label{eq95}
\end{align}
From eq. (\ref{eq95}) we get 
\begin{equation}
|\vec{x}_{\theta}|^{2}=\frac{{\xi_{x}}^{2}+\xi_{y}^{2}}{J^{2}}\cos^{2}\theta+\frac{{\eta_{x}}^{2}+{\eta_{y}}^{2}}{J^{2}}\sin^{2}\theta+2\frac{\xi_{x}\eta_{x}+\xi_{y}\eta_{y}}{J^{2}}\cos \theta \sin \theta
\label{eq96}
\end{equation}
Using the flow tangency condition $\phi_\eta \sin\theta + \phi_\xi \cos\theta = 0$ it can be shown that (\ref{eq96}) yields
\begin{equation}
|\vec{x}_{\theta}|=\frac{2\tilde{\rho}}{1+\tilde{\rho}}h
\label{eq97}
\end{equation}
From eq. (\ref{eq95}) we also get
\begin{equation}
\vec{x}_{\theta}\cdot\nabla\phi=|\vec{x}_{\theta}|\partial_{t}\phi=\phi_{x}x_{\theta}+\phi_{y}y_{\theta}=\phi_{\eta}\cos \theta-\phi_{\xi}\sin \theta=\partial_{t}\phi^{circle}
\label{eq98}
\end{equation}
Hence, 
\begin{equation}
\oint_{S}\vec{v}\cdot d\vec{x}=\oint_{S}\nabla\phi\cdot\vec{x}_{\theta}d\theta=\oint_{circle}\partial_{t}\phi d\theta
\label{eq99}
\end{equation}
and, thus, the parameter $\Gamma$ in eq. (\ref{eq84}), which is the circulation in the $\zeta$-plane, is also the circulation in the $z$-plane. 

Let us now consider a flow perturbation $\delta\vec{v}^{comp}(x,y)$ that can be written in terms of a perturbation potential $\delta\vec{v}^{comp}(x,y)=\nabla\delta\phi^{comp}(x,y)$. The perturbation causes a change in the mapping function. Since $\phi^{comp}(x,y)=\phi^{inc}(\xi(x,y),\eta(x,y))$, we have for the first variation
\begin{align}
\delta\phi^{comp}(x,y)&=\delta\phi^{inc}(\xi(x,y),\eta(x,y))+(\delta\xi,\delta\eta)\cdot\nabla\phi^{inc}(\xi(x,y),\eta(x,y))\\ \nonumber &= \delta\phi^{inc}+(\delta\xi,\delta\eta)\cdot\vec{v}^{inc}
\label{eq100}
\end{align}
Hence,
\begin{equation}
|\vec{x}_{\theta}|\hat{t}\cdot\delta\vec{v}^{comp}=|\vec{x}_{\theta}|\partial_{t}\delta\phi^{comp}(x,y)=\partial_{t}\delta\phi^{inc}+\partial_{t}(\delta\xi,\delta\eta)\cdot\vec{v}^{inc}+(\delta\xi,\delta\eta)\cdot\partial_{t}\vec{v}^{inc}
\label{eq101}
\end{equation}
The terms on the RHS of eq. (\ref{eq101}) are computed in the circle plane. At the rear stagnation point of the circle, the velocity vanishes, while for the normalized Bers mapping (which takes $(x,y)_{te}\rightarrow(1,0)$) $(\delta\xi,\delta\eta)_{rsp}=0$. Hence,
\begin{equation}
|\vec{x}_{\theta}|\hat{t}\cdot\delta\vec{v}_{te}^{comp}=|\vec{x}_{\theta}|\partial_{t}\delta\phi^{comp}(x,y)|_{te}=\partial_{t}\delta\phi^{inc}|_{rsp}
\label{eq102}
\end{equation}
which, using eq. (\ref{eq97}) yields eq. (\ref{eq47}).

\subsection{The circulation mode}


In \eqref{eqVarPot} we introduced the compressible unit-circulation mode $\phi^{(\Gamma)}$ in the airfoil plane. In the 2D incompressible case, this mode has a closed form solution in terms of a single layer potential \cite{coclici1999,bassanini1999}, while in the compressible setting it has to be understood as the derivative, with respect to circulation, of the family of subcritical potential flows with fixed geometry and free-stream data. The use of $\Gamma$ as the scalar circulation parameter is standard in the subsonic potential-flow problem \cite{finn1958}; under the normalized Bers correspondence \cite{bers1954}, this
mode is represented in the circle plane by the usual logarithmic circulation term
\begin{equation}
\frac{\Gamma}{2\pi i}\log\zeta .    
\end{equation}

Accordingly, the image of $\phi^{(\Gamma)}$ in the circle plane is the standard incompressible
unit-circulation mode, while the physical-plane mode is its quasi-conformal pullback, including
the linearized dependence of the mapping on the flow. In particular, the trailing-edge relation
\eqref{eq47}--\eqref{eq48}  gives, for the pure circulation mode,
\begin{equation}
\partial_t\phi^{inc,(\Gamma)}\big|_{rsp}=\frac{1}{2\pi},
\qquad
h\,\hat t\cdot\nabla\phi^{(\Gamma)}
\to
\frac{1+\tilde\rho_{te}}{4\pi\tilde\rho_{te}},    
\end{equation}
so that $\hat t\cdot\nabla\phi^{(\Gamma)}$ has the same trailing-edge exponent
$r^{-\beta}$ as in the incompressible case, but with a compressibility-dependent prefactor.

In subcritical potential flows, circulation $\Gamma$ is a free parameter prior to the application of the Kutta condition. Consequently, for given geometry and flow conditions there exists a family of exact full potential solutions continuously parameterized by the circulation. Differentiation with respect to $\Gamma$ yields a sensitivity mode that inherently satisfies the full-potential equation linearized about the baseline flow. Indeed, let \(\phi(\Gamma)\) denote the one-parameter family of subcritical potential-flow
solutions with fixed geometry and free-stream data, parametrized by the circulation. Each
member of the family satisfies
\begin{equation}
\nabla\cdot\left(\rho(\nabla\phi)\nabla\phi\right)=0 .
\end{equation}
Differentiating this equation with respect to \(\Gamma\) at the base solution gives
\begin{equation}
\nabla\cdot\left(\delta\rho\,\nabla\phi+\rho\nabla\phi^{(\Gamma)}\right)=0,
\qquad
\delta\rho=-\rho a^{-2}\nabla\phi\cdot\nabla\phi^{(\Gamma)} ,
\end{equation}
and therefore
\begin{equation}
\nabla\cdot\left(\hat A\nabla\phi^{(\Gamma)}\right)=0,
\end{equation}
with \(\hat A\) defined in \eqref{eqAhat}. Thus, \(\phi^{(\Gamma)}\) is a homogeneous solution of
the same linearized full-potential operator as any admissible potential perturbation, but with
unit circulation around the profile.

Of course, when the Kutta condition is imposed, this degeneracy is removed: the circulation is fixed, and the circulatory tangent mode generally does not satisfy the linearized Kutta condition.

\subsection{ Kutta condition}

We end by recalling a few facts about the behavior of the potential and its derivatives at sharp trailing edges. This is merely a restatement of Bers' results \cite{bers1954}. Let us suppose that the airfoil has a trailing edge with wedge angle $\tau$, $0\le\tau\le\pi$, the lower limit corresponding to a cusp and the upper limit to a smooth (rounded) trailing edge. We then have
\begin{equation}
|\zeta-1|\sim|z-z_{te}|^{\frac{1}{2-\tau/\pi}}
\label{eq103}
\end{equation}
and, thus,
\begin{equation}
h^{-1}=\left|\frac{\partial\zeta}{\partial z}\right|\sim|z-z_{te}|^{\frac{\tau/\pi-1}{2-\tau/\pi}}=|z-z_{te}|^{-\beta}
\label{eq104}
\end{equation}
where $\beta=(\pi-\tau)/(2\pi-\tau)$. Hence, the complex velocity $w(z)=\phi_{x}-i\phi_{y}$ (\ref{eq88}) behaves as
\begin{equation}
|w(z)|\sim|z-z_{te}|^{\frac{\tau/\pi-1}{2-\tau/\pi}}\left|A\left(e^{-i\alpha}-\frac{e^{i\alpha}}{\zeta^{2}}\right)+\frac{\Gamma}{2\pi i\zeta}\right|_{\zeta\rightarrow 1}=|z-z_{te}|^{\frac{\tau/\pi-1}{2-\tau/\pi}}\left|-2iA \sin \alpha+\frac{\Gamma}{2\pi i}\right|
\label{eq105}
\end{equation}
which is singular (for $\tau\ne\pi$) unless the bracketed term vanishes, i.e.,
\begin{equation}
\Gamma=-4\pi A \sin \alpha
\label{eq106}
\end{equation}
which is the Kutta condition, in which case the complex velocity
\begin{equation}
|w(z)|\sim|z-z_{te}|^{\frac{\tau/\pi-1}{2-\tau/\pi}}\left|A\left(e^{-i\alpha}-\frac{e^{i\alpha}}{\zeta^{2}}\right)+\frac{-4\pi A \sin \alpha}{2\pi i\zeta}\right|_{\zeta\rightarrow 1}\sim|z-z_{te}|^{\frac{\tau/\pi}{2-\tau/\pi}}
\label{eq107}
\end{equation}
is singularity-free and yields a stagnation point at the trailing edge whenever $\tau>0$.

\end{appendices}

\section*{Data Availability}
The data that support the findings of this study are available from the corresponding authors upon reasonable request.

\section*{Funding}
The research described in this paper has been supported by INTA under grant IDATEC (IGB21001).

\section*{Acknowledgments}
The numerical computations reported in the paper have been carried out with the SU2 code, an open source platform developed and maintained by the SU2 Foundation.

\end{document}